\documentclass[aip,amsmath,amssymb,reprint]{revtex4-2}

\usepackage{graphicx}
\usepackage{dcolumn}
\usepackage{bm}
\usepackage[modulo]{lineno}

\usepackage{makecell}
\usepackage{booktabs,tabularx}
\usepackage[utf8]{inputenc}
\usepackage[T1]{fontenc}
\usepackage{mathptmx}
\usepackage{etoolbox}
\usepackage{xcolor}
\usepackage{hyperref}
\hypersetup{%
	pdfpagemode={UseOutlines},
	bookmarksopen,
	pdfstartview={FitH},
	colorlinks,
	linkcolor={black},
	citecolor={black},
	urlcolor={black}
}
\usepackage{parskip}
\usepackage{braket}
\usepackage{multirow}
\usepackage{indentfirst}
\usepackage{parskip}
\makeatletter
\def\@email#1#2{%
 \endgroup
 \patchcmd{\titleblock@produce}
  {\frontmatter@RRAPformat}
  {\frontmatter@RRAPformat{\produce@RRAP{*#1\href{mailto:#2}{#2}}}\frontmatter@RRAPformat}
  {}{}
}%
\makeatother

\newcommand*\rot{\rotatebox{90}}

\begin{document}
	
\title{Superconducting microwave oscillators as detectors for ESR spectroscopy}

\author{R. Russo}
\email{roberto.russo@epfl.ch}
\affiliation{ 
	Microsystems Laboratory, Ecole Polytechnique Fédérale de Lausanne (EPFL), 1015, Lausanne, Switzerland 
}
\affiliation{ 
	Center for Quantum Science and Engineering, Ecole Polytechnique Fédérale de Lausanne (EPFL), 1015, Lausanne, Switzerland }

\author{A. Chatel}

\affiliation{ 
	Microsystems Laboratory, Ecole Polytechnique Fédérale de Lausanne (EPFL), 1015, Lausanne, Switzerland 
}
\affiliation{ 
	Center for Quantum Science and Engineering, Ecole Polytechnique Fédérale de Lausanne (EPFL), 1015, Lausanne, Switzerland }
\author{N. Brusadin}
\affiliation{ 
	Microsystems Laboratory, Ecole Polytechnique Fédérale de Lausanne (EPFL), 1015, Lausanne, Switzerland 
}
\author{R. Yu}
\affiliation{ 
	Microsystems Laboratory, Ecole Polytechnique Fédérale de Lausanne (EPFL), 1015, Lausanne, Switzerland 
}
\author{R. Farsi}
\affiliation{ 
	Microsystems Laboratory, Ecole Polytechnique Fédérale de Lausanne (EPFL), 1015, Lausanne, Switzerland 
}

\author{H. Furci}
\affiliation{ 
	Microsystems Laboratory, Ecole Polytechnique Fédérale de Lausanne (EPFL), 1015, Lausanne, Switzerland 
}

\author{J. Brugger}
\affiliation{ 
	Microsystems Laboratory, Ecole Polytechnique Fédérale de Lausanne (EPFL), 1015, Lausanne, Switzerland 
}

\author{G. Boero}
\affiliation{ 
	Microsystems Laboratory, Ecole Polytechnique Fédérale de Lausanne (EPFL), 1015, Lausanne, Switzerland 
}
\affiliation{ 
	Center for Quantum Science and Engineering, Ecole Polytechnique Fédérale de Lausanne (EPFL), 1015, Lausanne, Switzerland }

\date{\today}

\begin{abstract}
Microwave superconducting resonators are extensively studied in fields such as quantum computing and electron spin resonance (ESR) spectroscopy. However, the integration of superconducting resonators with feedback mechanisms to create ultra-low noise oscillators is a relatively unexplored area, and the application of such oscillators in ESR spectroscopy has not yet been demonstrated. In this work, we report the design, fabrication, and application of microwave oscillators based on superconducting resonators for ESR spectroscopy, illustrating an alternative way for the improvement of the performance of oscillator based ESR sensors. Specifically, ESR spectra are obtained by measuring the oscillator's frequency shift induced by the ESR effect as a function of the applied static magnetic field. The oscillators are composed of a single heterojunction bipolar transistor (HBT) or high electron mobility transistor (HEMT) coupled with NbTi or YBa$_2$Cu$_3$O$_7$ (YBCO) superconducting resonators. The fabricated oscillators operate at frequencies of 0.6 and 1.7 GHz and temperatures up to 80 K (for YBCO resonators) and 8 K (for NbTi resonators). The lowest measured frequency noise is about 9 mHz/Hz$^{1/2}$ (-139 dBc/Hz), the best spin sensitivity is about $1\times10^{10}$ spins/Hz$^{1/2}$, and the best concentration sensitivity is about $3\times10^{18}$ spins/Hz$^{1/2}$m$^3$. The approach proposed in this work should allow for significantly better spin and concentration sensitivities compared to those achievable with normal conductors, up to operating frequencies, magnetic fields, and temperatures where superconductors exhibit substantially lower effective microwave resistance than normal conductors.
\end{abstract}

\maketitle

Planar superconducting resonators are widely recognized as pivotal empowering tools both in research studies for quantum computing applications \cite{Frunzio2005,Goppl2008,samkharadze2016,ghirri2015,ghirri2016,bonizzoni2016} and for the improvements brought to the field of ESR spectroscopy \cite{bienfait2016,artzi2022,dayan2018,asfaw2017,probst2017,Morton2018,Abhyankar2022,Blank2017,bonizzoni2023,Malissa2013,Golovchanskiy2018,Scheffler2013}. For ESR spectroscopy at low temperature, miniaturized superconducting resonators represent a promising approach to study subnanoliter samples with sensitivities down to a few spins/Hz$^{1/2}$ at milli-Kelvin temperatures. Such resonators have been realized using superconducting materials such as YBa$_2$Cu$_3$O$_7$ (YBCO) \cite{bonizzoni2016,artzi2022,ghirri2015,dayan2018}, Nb \cite{kwon2018,rausch2018,zhang2016}, NbN \cite{Carter2019,mahashabde2020}, NbTi \cite{Russo2023},NbTiN \cite{kroll2019,asfaw2017,samkharadze2016}, and Al \cite{probst2017}.

In the 80s, Stein and Turneaure \cite{stein1978} and Stein \cite{stein1980} underlined how the "tremendous stability potential of superconducting resonators" could lead to realize ultra low noise oscillators. Komiyama et al. \cite{komiyama1987}, following an idea discussed by Stein \cite{stein1978}, reported on the realization of a 9.3 GHz Nb cavity stabilized oscillator. Borghs et al. \cite{borghs1994} presented the integration of an YBCO resonator with a single HEMT, obtaining a 12 GHz oscillator with a frequency noise of 2.5 Hz/Hz$^{1/2}$ at 10 kHz (-72 dBc/Hz). Some years later, Kojouharov et al. \cite{kojouharov1997} demonstrated a  12 GHz microwave microstrip oscillator stabilized with a Tl$_2$Ba$_2$CaCu$_2$O$_8$ resonator achieving 1.8 Hz/Hz$^{1/2}$ at 100 kHz (-95 dBc/Hz). Ghosh et al. \cite{ghosh1997} showed a LaAlO$_3$ high temperature supercondutor (HTS) shielded dielectric resonator working at 5.6 GHz. Placed in feedback with a room-temperature amplifier, this configuration resulted in a frequency noise of 4.5 mHz/Hz$^{1/2}$ at 10 kHz (-127 dBc/Hz). Recently, Chaudy et al. \cite{Chaudy2020,chaudy2017} demonstrated a superconducting oscillator operating at 1 GHz realized by the combination of an YBCO resonator on MgO and of a SiGe heterojunction bipolar transistor (HBT), resulting in an exceptionally low frequency noise of about 300 $\mu$Hz/Hz$^{1/2}$ at 100 kHz (-170 dBc/Hz) and 65 K. This achievement stimulated our interest in the realization of low noise oscillators based on superconducting resonators, specifically conceived for ESR spectroscopy. 

\begin{figure*}[hht!] 
	\includegraphics[width=\textwidth]{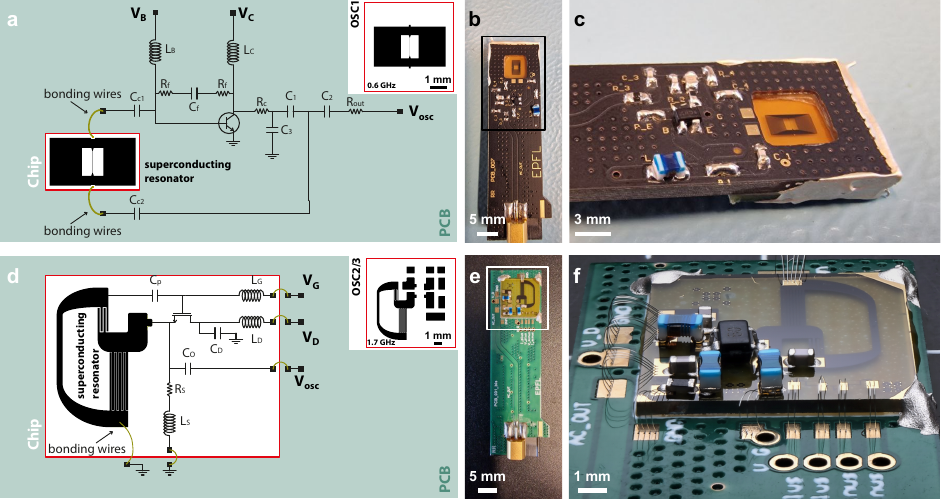} 
	\caption{Structure and assembly of the oscillators. \textbf{a}, Circuit of the 0.6 GHz SiGe HBT based oscillator (OSC1, YBCO). Bonding wires are used to connect the resonator to the feedback electronics. $V_B$: base bias voltage, $V_C$: collector bias voltage, $V_{OSC}$: oscillator output signal. The resonator is realized on a Al$_2$O$_3$ chip, whereas all other electronics components of the feedback are soldered on a printed circuit board (PCB). \textbf{b}, Photo of the PCB. \textbf{c}, Details of the oscillator. \textbf{d}, Circuit for the pHEMT based oscillators operating at 1.73 GHz (OSC2, YBCO) and 1.72 GHz (OSC3, NbTi). $V_G$: gate bias voltage, $V_D$: drain bias voltage, $V_{OSC}$: oscillator output signal \textbf{e}, Photo of the PCB. \textbf{f}, Details of the oscillators. All components are directly glued on the Al$_2$O$_3$ chip together with the resonator.}
	\label{figure: circuit_and_assembly}
	\makeatletter
	\let\save@currentlabel\@currentlabel
	\edef\@currentlabel{\save@currentlabel a}\label{figure: circuit_and_assembly:a}
	\edef\@currentlabel{\save@currentlabel b}\label{figure: circuit_and_assembly:b}
	\edef\@currentlabel{\save@currentlabel c}\label{figure: circuit_and_assembly:c}
	\edef\@currentlabel{\save@currentlabel d}\label{figure: circuit_and_assembly:d}
	\edef\@currentlabel{\save@currentlabel e}\label{figure: circuit_and_assembly:e}
	\edef\@currentlabel{\save@currentlabel f}\label{figure: circuit_and_assembly:f}
	\edef\@currentlabel{\save@currentlabel a-1c}\label{figure: circuit_and_assembly:a-c}
	\edef\@currentlabel{\save@currentlabel d-1f}\label{figure: circuit_and_assembly:d-f}
	\makeatother
\end{figure*}

Sample-induced frequency shift in a microwave LC oscillator under varying static magnetic fields has been extensively studied to perform ESR spectroscopy \cite{Yalcin2008,anders2012,gualco2014,matheoud2018,Schlecker2019,Hassan2021,Chu2023,Kunstner2024,kern2024,Segantini2024,matheoud2017}. Such LC oscillators comprise a conductor-based resonator and transistor-based feedback electronics. In this letter, we report on the realization and characterization of microwave LC oscillators for ESR spectroscopy
based on superconducting resonators. In particular, we investigated two superconducting materials, YBCO and NbTi, and two transistors, heterojunction bipolar transistors (HBT) and high electron mobility transistors (HEMT). Superconducting resonators can achieve significantly larger $Q$-factors than conducting resonators, at least up to several GHz \cite{Blank2017,bonizzoni2016,artzi2022,ghirri2015,dayan2018,zhang2016,asfaw2017,probst2017}. Hence, their use should allow to reduce the oscillator phase noise (PN) and improve the spin sensitivity. In this work, we demonstrate that it is indeed possible to reduce the oscillator PN and improve the spin sensitivity but we underline also that such improvements are strongly influenced by the ESR experimental conditions.   

The YBCO resonators are realized starting from chips with a YBCO(300 nm)/CeO$_2$(20 nm) layer on a single crystal Al$_2$O$_3$ (sapphire) substrate (Ceraco GmBH). The NbTi resonators are realized starting from 4-inches sapphire wafers (Siegert wafer GmbH) on which we sputtered a 150 nm layer of NbTi. Details on the fabrication of the YBCO and NbTi resonators are reported in the \textit{supplementary material}.

\begin{figure*}[ht!]
	\centering 
	\includegraphics[width=\linewidth]{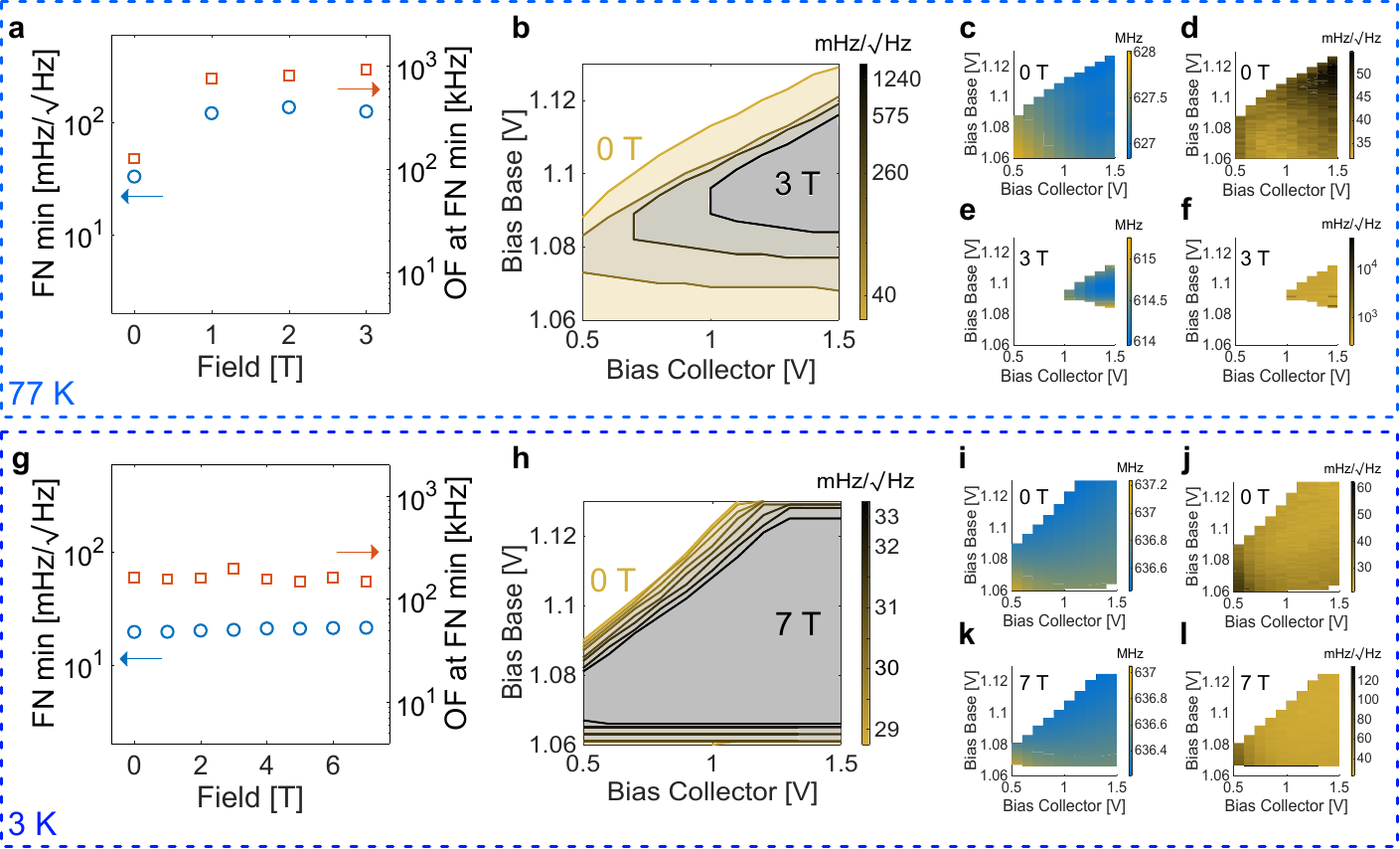}
	\caption{0.6 GHz oscillator (OSC1) frequency and frequency noise (FN) vs bias voltages, magnetic field, and temperature. The frequency noise (FN, in Hz/Hz$^{1/2}$) is calculated from the measured phase noise (PN, in dBc/Hz) as FN$=$OF$\times10^{\rm{PN}/20}$, where OF is the offset frequency (in Hz). 
		\textbf{a}, \textbf{g}, Minimum FN and corresponding OF vs magnetic field respectively at 77 K and 3 K. 
		\textbf{b}, \textbf{h}, Bias voltages working region and FN vs magnetic field respectively at 77 K and 3 K (reported FN values are the average value at 100 kHz OF in the corresponding biasing region).
		\textbf{c}, \textbf{e}, \textbf{i}, \textbf{k}, Oscillation frequency vs bias voltages at 0 and 3 T / 7 T, both at 77 K (\textbf{c}, \textbf{e}) and 3 K (\textbf{i}, \textbf{k}). 
		\textbf{d}, \textbf{f}, \textbf{j}, \textbf{l}, FN at 100 kHz OF vs bias voltages at 0 and 3 T / 7 T, both at 77 K (\textbf{d}, \textbf{f}) and 3 K (\textbf{j}, \textbf{l}).} 	
	\label{figure: working_area__freq__noise}
	\makeatletter
	\let\save@currentlabel\@currentlabel
	\edef\@currentlabel{\save@currentlabel a}\label{figure: working_area__freq__noise:a}
	\edef\@currentlabel{\save@currentlabel b}\label{figure: working_area__freq__noise:b}
	\edef\@currentlabel{\save@currentlabel c}\label{figure: working_area__freq__noise:c}
	\edef\@currentlabel{\save@currentlabel d}\label{figure: working_area__freq__noise:d}
	\edef\@currentlabel{\save@currentlabel e}\label{figure: working_area__freq__noise:e}
	\edef\@currentlabel{\save@currentlabel f}\label{figure: working_area__freq__noise:f}
	\edef\@currentlabel{\save@currentlabel g}\label{figure: working_area__freq__noise:g}
	\edef\@currentlabel{\save@currentlabel h}\label{figure: working_area__freq__noise:h}
	\edef\@currentlabel{\save@currentlabel i}\label{figure: working_area__freq__noise:i}
	\edef\@currentlabel{\save@currentlabel j}\label{figure: working_area__freq__noise:j}
	\edef\@currentlabel{\save@currentlabel k}\label{figure: working_area__freq__noise:k}
	\edef\@currentlabel{\save@currentlabel l}\label{figure: working_area__freq__noise:l}
	\makeatother
\end{figure*}

To realize the oscillators, the resonators are connected to feedback electronic circuitry. In this study, two different approaches are implemented. The former is rooted in what presented by Chaudy et al. \cite{Chaudy2020}. The oscillator is built by combining three elements: a superconducting resonator, working as a bandpass filter and fixing the oscillation frequency, an RF amplifier whose bandwidth includes the resonator frequency and few circuital elements to adjust the phase and respect Barkhausen conditions for oscillation. This approach is later referred as "blocks" approach. The latter adopts a Colpitts configuration, as reported for a non superconducting integrated oscillator by Matheoud et al. \cite{Matheoud2019_2}. Overall, three oscillators are realized: oscillator 1 (OSC1), oscillator 2 (OSC2) and oscillator 3 (OSC3) (Fig.\ref{figure: circuit_and_assembly}). OSC1 uses the blocks approach. In this case, a low noise RF SiGe HBT transistor (BFP650, Infineon) is used. The amplifier consists of a common emitter stage, with the resonator connected between base and collector. To grant flexibility, base and collector are biased independently. Two capacitors C$_{c1}$ and C$_{c2}$ are added on the feedback loop, on the resonator's sides, to adjust the phase and the coupling with the resonator. The resonator is realized on a 10$\times$10 mm$^{2}$ chip (see Fig.\ref{figure: circuit_and_assembly:a-c}). OSC2 and OSC3 have a Colpitts configuration, characterized by a superconducting LC resonator with a capacitive feedback. Gate and drain are, again, biased independently. This requires a capacitor to decouple the superconducting resonator from the gate DC biasing, avoiding DC current flowing in the resonator. In this approach, the entire feedback electronics is mounted onto the 10$\times$10 mm$^{2}$ chip together with the fabricated resonator. The electronic components are glued on the chip's pads (see Fig.\ref{figure: circuit_and_assembly:d-f}). For components values and assembly details on all oscillators see the \textit{supplementary material}.    

The electrical characterization of the oscillators is performed in liquid nitrogen (LN$_2$, 77 K) or liquid helium (LHe, 4.2 K) dewars as well as in the variable temperature insert of a superconducting magnet (0 to 9.4 T, 1.4 to 325 K, Cryogenic Ltd). The oscillation frequency and frequency noise of the oscillators are measured using a phase noise analyzer (Agilent Technologies, E5052A).  

The ESR experiments are performed in the same superconducting magnet, with the oscillators in static He gas. All ESR experiments are performed with crystals of $\alpha$,$\gamma$-bisdiphenylen-$\beta$-phenylallyl (BDPA, Sigma-Aldrich 152560). The oscillators microwave output signal is routed outside the magnet and downconverted to about 10 MHz using a frequency mixer. The 10 MHz signal is delivered to a custom phase-locked-loop (PLL) acting as frequency-to-voltage converter. The converted signal is directed to a lock-in amplifier performing the synchronous demodulation at the frequency of the magnetic field modulation. For a detailed description of the set-up used for ESR measurements see the \textit{supplementary material}.

This set-up has shown some limitations: the frequency noise measured on the ESR spectra (i.e., with the oscillator in the superconducting magnet and using the electronics described above) is up to 100 times higher than the intrinsic frequency noise of the oscillators (defined as the frequency noise measured with the oscillator placed in dewars of LN$_2$ or LHe). 

Each oscillator shows a different frequency noise degradation due to the experimental set-up, with OSC1 being the less influenced. The origin of the frequency noise increase is unclear (see discussion in the \textit{supplementary material}).

\begin{figure*}[ht!]
	\centering 
	\includegraphics[width=\linewidth]{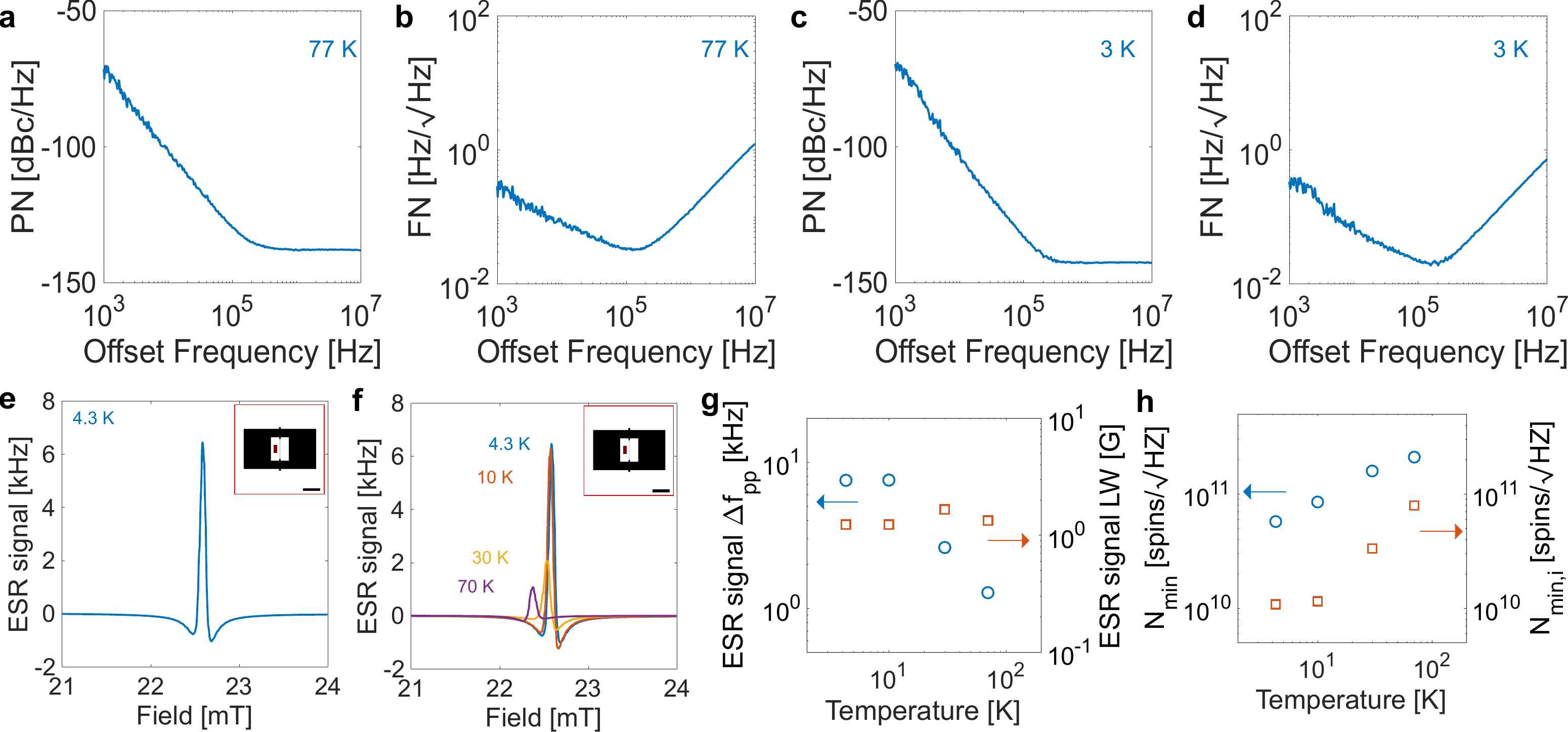}
	\caption{Experiments with the 0.6 GHz superconducting oscillator (OSC1). All ESR spectra are measured with a BDPA sample (volume of 400$\times$150$\times$20 $\mu$m$^3$) and with a field modulation amplitude of 0.4 G. \textbf{a}, PN spectrum at 77 K, V$_{B}$ = 1.06 V, V$_{C}$ = 1 V. \textbf{b}, FN spectrum corresponding to \textbf{a}. \textbf{c}, PN spectrum at 3 K, V$_{B}$ = 1.13 V, V$_{C}$ = 1.5 V. \textbf{d}, FN spectrum corresponding to \textbf{c}. \textbf{e}, ESR spectrum at 4.3 K. The inset indicates the BDPA sample position within the resonator sensitive volume (scale bar 1 mm). \textbf{f}, ESR spectra at different temperatures. \textbf{g}, ESR signal peak-to-peak amplitude and linewidth vs temperature. \textbf{h}, Spin sensitivity vs temperature, considering  the FN measured on the ESR spectra and the intrinsic FN.}
	\label{figure: ESR_600_MHz_s1}
	\makeatletter
	\let\save@currentlabel\@currentlabel
	\edef\@currentlabel{\save@currentlabel a}\label{figure: ESR_600_MHz_s1:a}
	\edef\@currentlabel{\save@currentlabel b}\label{figure: ESR_600_MHz_s1:b}
	\edef\@currentlabel{\save@currentlabel c}\label{figure: ESR_600_MHz_s1:c}
	\edef\@currentlabel{\save@currentlabel d}\label{figure: ESR_600_MHz_s1:d}
	\edef\@currentlabel{\save@currentlabel e}\label{figure: ESR_600_MHz_s1:e}
	\edef\@currentlabel{\save@currentlabel f}\label{figure: ESR_600_MHz_s1:f}
	\edef\@currentlabel{\save@currentlabel g}\label{figure: ESR_600_MHz_s1:g}
	\edef\@currentlabel{\save@currentlabel h}\label{figure: ESR_600_MHz_s1:h}
	\makeatother
\end{figure*}

\begin{figure*}[ht!]
	\centering 
	\includegraphics[width=\linewidth]{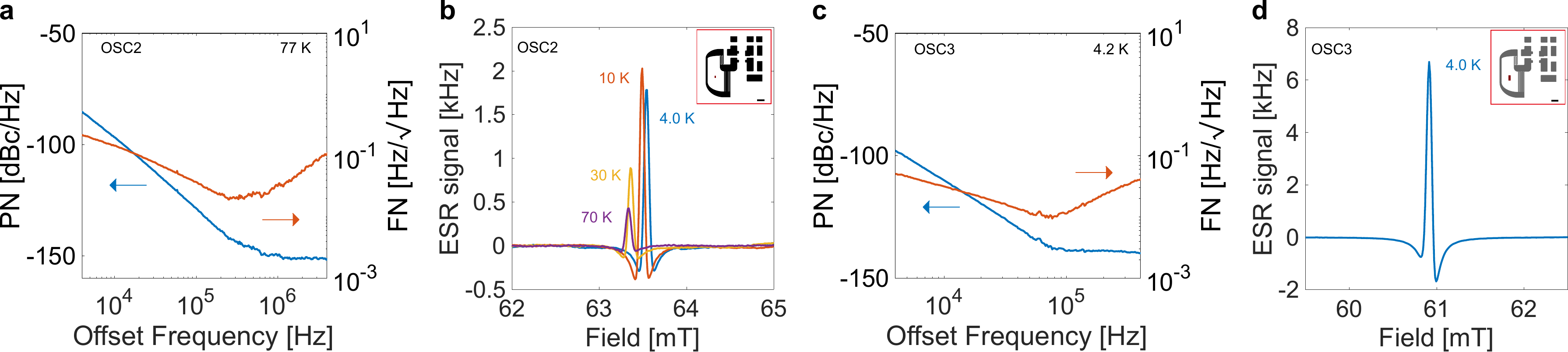} 
	\caption{Experiments with the 1.7 GHz superconducting oscillators (OSC2 and OSC3).
		\textbf{a}, PN and FN of OSC2 at 77 K with V$_{G}$ =-0.66 V and V$_{D}$ = 1.12 V. \textbf{b}, ESR spectra acquired with OSC2 at 4 K, 10 K, 30 K and 70 K on a BDPA sample (volume of 300$\times$200$\times$50 $\mu$m$^3$) with a field modulation amplitude of 0.4 G. The inset indicates the BDPA sample position within the resonator sensitive volume (scale bar 1 mm) \textbf{c}, PN and and FN of OSC3 at 4.2 K with V$_{G}$ =-0.71 V and V$_{D}$ = 0.7 V. \textbf{d}, ESR spectrum acquired with OSC3 at 4 K on a BDPA sample (volume of 750$\times$300$\times$50 $\mu$m$^3$) with a field modulation amplitude of 0.4 G. The inset indicates the BDPA sample position within the resonator sensitive volume (scale bar 1 mm)}
	\label{figure: composition_ESR_all_OSC_1_2_4_5_v1}
	\makeatletter
	\let\save@currentlabel\@currentlabel
	\edef\@currentlabel{\save@currentlabel a}\label{figure: composition_ESR_all_OSC_1_2_4_5_v1:a}
	\edef\@currentlabel{\save@currentlabel b}\label{figure: composition_ESR_all_OSC_1_2_4_5_v1:b}
	\edef\@currentlabel{\save@currentlabel c}\label{figure: composition_ESR_all_OSC_1_2_4_5_v1:c}
	\edef\@currentlabel{\save@currentlabel d}\label{figure: composition_ESR_all_OSC_1_2_4_5_v1:d}
	\makeatother
\end{figure*}

In Fig.\ref{figure: working_area__freq__noise} we report a detailed analysis of OSC1 (0.6 GHz). Oscillation frequency and frequency noise are investigated at different voltage bias points, static magnetic fields, and temperatures. The static magnetic field is applied parallel to the resonator plane. At 77 K, the magnetic field is swept from 0 to 3 T. The biasing range for oscillation shrinks with the magnetic field (see Fig.\ref{figure: working_area__freq__noise:b}). At fields higher than 3 T, the resonator $Q$-factor becomes too low for oscillation. At 3 K the field is applied from 0 to 7 T. At this temperature, well below YBCO $T_c$ (about 90 K), the $Q$-factor of the resonator is marginally influenced by the magnetic field and the oscillator operates up to at least 7 T (see Fig.\ref{figure: working_area__freq__noise:h}). The biasing range is also not significantly influenced by the magnetic field. At 77 K, the oscillator frequency noise increases significantly with the magnetic field, as a consequence of the degradation of the resonator $Q$-factor. At 77 K, the frequency noise spectrum shows a minimum of 30 mHz/Hz$^{1/2}$ (-132 dBc/Hz) at about 100 kHz from the carrier at 0 T and of about 125 mHz/Hz$^{1/2}$ (-137 dBc/Hz) at about 1 MHz from the carrier at 3 T (see Fig.\ref{figure: working_area__freq__noise:a}). At 3 K, the frequency noise is only marginally perturbed by static magnetic fields up to at least 7 T (see Fig.\ref{figure: working_area__freq__noise:g}). At 3 K and 200 kHz from carrier, the frequency noise spectrum shows a minimum of 18 mHz/Hz$^{1/2}$ (-140 dBc/Hz). This frequency noise reduction of approximately a third from 77 K to 3 K is directly correlated to the $Q$-factor increase with reducing temperature, from about 6000 at 77 K to about 10000 at 3 K. Additional results on OSC1 are reported in the \textit{supplementary material}.

The ESR experiments with OSC1 are performed using a crystal of BDPA having a volume of 400$\times$150$\times$20 $\mu$m$^3$ placed on the resonator as in Fig.\ref{figure: ESR_600_MHz_s1:e}. The ESR signal linewidth is about 1.2 G, as expected with a $B_1\cong$ 0.2 G (see $B_1$ estimation in the \textit{supplementary material}). At 4.3 K, a spin sensitivity $N_{min}\cong6\times$10$^{10}$ spins/Hz$^{1/2}$ and a concentration sensitivity $C_{min}\cong 9\times$10$^{19}$ spins/Hz$^{1/2}$m$^3$ are obtained. As mentioned above, these values are affected by an experimental set-up induced frequency noise degradation. Indeed, considering the oscillator intrinsic noise, a spin sensitivity $N_{min,i}\cong$ 1$\times$10$^{10}$ spins/Hz$^{1/2}$ and a concentration sensitivity $C_{min,i}\cong$ 2$\times$10$^{19}$ spins/Hz$^{1/2}$m$^3$ could be reached (see Fig.\ref{figure: ESR_600_MHz_s1:h}). ESR spectra acquired at different temperatures (see Fig.\ref{figure: ESR_600_MHz_s1:f}) show that the signal amplitude decreases respecting Curie's law from 10 to 70 K (see Fig.\ref{figure: ESR_600_MHz_s1:g}).

\begin{table*}[!ht]
	\begin{tabular}{rrrrrrrrrrrrrrrrr} 
		\rot{\thead{\text{Device}}} & \rot{\thead{\text{Resonator} \\ \text{Material}}} & \rot{\thead{\text{Active} \\ \text{Device}}} & \rot{\thead{\text{Frequency} \\ \text{[GHz]}}} & \rot{\thead{\text{Temperature} \\ \text{[K]}}} & \rot{\thead{\text{OF$_{min}$}  \text{[kHz]}}} &  \rot{\thead{\text{FN @ O$F_{min}$} \\ \text{[mHz/Hz$^{1/2}$]}}} & \rot{\thead{\text{FN @ 10 kHz} \\ \text{[mHz/Hz$^{1/2}$]}}} & \rot{\thead{\text{FN @ 100 kHz} \\ \text{[mHz/Hz$^{1/2}$]}}} & \rot{\thead{\text{OF$_{ESR}$}  \text{[kHz]}}} & \rot{\thead{\text{FN$_{ESR}$} \\ \text{[mHz/Hz$^{1/2}$]}}} & \rot{\thead{\text{Power} \\ \text{consumption} \\ \text{[mW]}}}  & \rot{\thead{\text{N$_{min}$} \\ \text{[spins/Hz$^{1/2}$]}\\ \\ \\}}  & \rot{\thead{\text{Effective}  \text{volume} \\ \text{[mm$^{3}$]}}}  & \rot{\thead{\text{C$_{min}$} \\ \text{[spins/Hz$^{1/2}$m$^3$]}\\ \\}}  \\ \hline \hline 
		\thead{\text{OSC1}}	& \thead{\text{YBCO}} & \thead{\text{SiGe} \\  \text{HBT}} & \thead{0.63} & \thead{\text{4.3} \\ \\ \text{10} \\ \\ \text{30} \\ \\ \text{70} \\ \\  \text{77*}}  & \thead{\text{25} \\ \\ \text{23} \\ \\ \text{27} \\ \\ \text{29} \\ \\  \text{48}} & \thead{\text{15} \\ \\ \text{16} \\ \\ \text{16} \\ \\ \text{19}  \\ \\ \text{24}} & \thead{\text{23} \\ \\ \text{25} \\ \\ \text{23} \\ \\ \text{28} \\ \\ \text{34}} & \thead{\text{35} \\ \\ \text{41} \\ \\ \text{34} \\ \\ \text{37} \\ \\ \text{30}} & \thead{\text{22} \\ \\ \text{22} \\ \\ \text{20} \\ \\ \text{20} \\ \\ \text{-}}& \thead{\text{79} \\ \\ \text{118} \\ \\ \text{76} \\ \\ \text{50} \\ \\ \text{-}}	 & \thead{\text{21} \\ \\ \text{21} \\ \\ \text{18} \\ \\ \text{11}  \\ \\ \text{6}} & \thead{ \text{6$\times$10$^{10}$}	\\ \\ \text{9$\times$10$^{10}$} \\ \\ \text{2$\times$10$^{11}$}\\ \\ \text{2$\times$10$^{11}$} \\ \\ \text{(9$\times$10$^{10}$)}}	 &  \thead{\text{0.7}} & \thead{ \text{9$\times$10$^{19}$}  \\ \\ \text{1$\times$10$^{20}$} \\ \\ \text{2$\times$10$^{20}$}  \\ \\ \text{3$\times$10$^{20}$} \\ \\ \text{(1$\times$10$^{20}$)}} \\	\hline 
		\thead{\text{OSC2}} 	& \thead{\text{YBCO}} & \thead{\text{pHEMT}} & \thead{1.73} & \thead{\text{4} \\ \\ \text{10} \\ \\ \text{30} \\ \\ \text{70}\\ \\ \text{77*}} & \thead{\text{9823} \\ \\ \text{9311} \\ \\ \text{6998} \\ \\ \text{9649}\\ \\ \text{249}} & \thead{\text{85} \\ \\ \text{84} \\ \\ \text{89} \\ \\ \text{84}\\ \\ \text{19}} & \thead{\text{2140} \\ \\ \text{2220} \\ \\ \text{2446} \\ \\ \text{3686}\\ \\ \text{146}} & \thead{\text{725} \\ \\ \text{737} \\ \\ \text{772} \\ \\ \text{948}\\ \\ \text{37}} & \thead{\text{72} \\ \\ \text{72} \\ \\ \text{72} \\ \\ \text{72}\\ \\ \text{-}} & \thead{\text{1304} \\ \\ \text{2004} \\ \\ \text{836} \\ \\ \text{1582}\\ \\ \text{-}} & \thead{\text{2} \\ \\ \text{2} \\ \\ \text{2.1} \\ \\ \text{2.3}\\ \\ \text{6}} & \thead{\text{9$\times$10$^{12}$}  \\	\\ \text{1$\times$10$^{13}$}  \\ \\ \text{1$\times$10$^{13}$} \\ \\ \text{4$\times$10$^{13}$} \\ \\ \text{(5$\times$10$^{11}$)}} & \thead{\text{19}} & \thead{\text{5$\times$10$^{20}$} \\	\\ \text{6$\times$10$^{20}$}  \\ \\ \text{6$\times$10$^{20}$} \\ \\ \text{2$\times$10$^{21}$} \\ \\ \text{(3$\times$10$^{19}$)}}  \\ \hline
		\thead{\text{OSC3}} 	& \thead{\text{NbTi}} & \thead{\text{pHEMT}} & \thead{1.72} & \thead{\text{4} \\ \\ \text{4.2*}} & \thead{\text{3024} \\ \\ \text{77}}  & \thead{\text{68} \\ \\ \text{9}} & \thead{\text{586} \\ \\ \text{32}} & \thead{\text{220} \\ \\ \text{11}} & \thead{\text{72} \\ \\ \text{-}} & \thead{\text{970} \\ \\ \text{-}} & \thead{\text{3.8} \\ \\ \text{2.1}}   & \thead{\text{6$\times$10$^{12}$} \\ \\ \text{(5$\times$10$^{10}$)} }  & \thead{\text{19}}  & \thead{\text{3$\times$10$^{20}$} \\ \\ \text{(3$\times$10$^{18}$)}}  \\ \hline \hline 
	\end{tabular}
	 \caption{Key-features of the oscillators, including resonator and feedback electronics characteristics, oscillation frequency, minimum FN, power consumption, effective sensitive volume, spin and concentration sensitivity. The output power of OSC1 is of about -30 dBm, whereas the output power of both OSC2 and OSC3 is of about -5 dBm. The effective volume is estimated as the volume resulting from the product of the resonator surface (the rectangular area around the central inductive line in case of the resonator of OSC1 and the area enclosed by the rounded rectangular coil and the capacitors in the case of OSC2 and OSC3) multiplied by an height approximated with the half-side of the resonator surface. For additional details see the \textit{supplementary material.} The asterisk (*) on the measurement temperature indicates that the FN is measured in a dewar of LN$_2$ or LHe. The values of spin sensitivity and concentration sensitivity in parenthesis are obtained considering the ESR signal amplitude (in Hz) measured in the superconducting magnet and the minimum intrinsic FN (in Hz/Hz$^{1/2}$) measured in the dewar of LN$_2$ or LHe.} 
	\label{oscillators_results}
\end{table*}

In the following we report on the other two oscillators. They operate at higher frequencies (1.7 GHz instead of 0.6 GHz) and have larger dimensions and sensitive volumes (19 $\mu$L instead of 0.7 $\mu$L). Additionally, they are based on a different transistor (HEMT instead of HBT) and on different superconducting materials (YBCO for OSC2 and NbTi for OSC3).

Fig.\ref{figure: composition_ESR_all_OSC_1_2_4_5_v1} presents the results of frequency noise and ESR experiments performed with OSC2 and OSC3. The measurement setup is the same as for OSC1. 

OSC2, measured at 77 K in a LN$_{2}$ dewar, oscillates at 1.73 GHz and has a minimum frequency noise of 19 mHz/Hz$^{1/2}$ (-142 dBc/Hz) at 250 kHz (see Fig.\ref{figure: composition_ESR_all_OSC_1_2_4_5_v1:a}). The lowest frequency noise is achieved with OSC3.
At 4.2 K in a LHe dewar, this oscillator operates at 1.72 GHz and has a minimum frequency noise of 9 mHz/Hz$^{1/2}$ (-139 dBc/Hz) at 77 kHz (see Fig.\ref{figure: composition_ESR_all_OSC_1_2_4_5_v1:c}). 

In Fig.\ref{figure: composition_ESR_all_OSC_1_2_4_5_v1:b} are reported the spectra of a BDPA sample ($300 \times200\times50$ $\mu$m$^3$) placed in the center of OSC2 resonator. At 4 K, a spin sensitivity $N_{min}\cong 9\times10^{12}$ spins/Hz$^{1/2}$ and a concentration sensitivity $C_{min}\cong 5\times10^{20}$ spins/Hz$^{1/2}$m$^3$ are estimated from the spectrum. At 70 K, a spin sensitivity $N_{min}\cong 4\times10^{13}$ spins/Hz$^{1/2}$ and a concentration sensitivity $C_{min}\cong 2\times10^{21}$ spins/Hz$^{1/2}$m$^3$ are obtained. These values are affected by the noise degradation in the cryomagnet. Indeed, considering the oscillator intrinsic noise, at 70 K, a spin sensitivity $N_{min,i}\cong 5\times10^{11}$ spins/Hz$^{1/2}$ and a concentration sensitivity $C_{min,i}\cong 3\times10^{19}$ spins/Hz$^{1/2}$m$^3$ could be reached. The ESR experiments are performed at 4, 10, 30 and 70 K, showing no signal broadening, indicating that the microwave field $B_1$ is much lower than 1 G. As for OSC1, the signal amplitude decreases respecting Curie's law from 10 to 70 K. Finally, Fig.\ref{figure: composition_ESR_all_OSC_1_2_4_5_v1:d} shows the ESR spectrum of a BDPA sample ($750\times300\times50$ $\mu$m$^3$) measured with OSC3. In this last case, a spin sensitivity $N_{min}\cong 6\times10^{12}$ spins/Hz$^{1/2}$ and a concentration sensitivity $C_{min}\cong 3\times10^{20}$ spins/Hz$^{1/2}$m$^3$ are estimated from the ESR spectrum, with again an influence of the measurement environment. Indeed, considering again the oscillator intrinsic noise, a spin sensitivity $N_{min,i}\cong 5\times10^{10}$ spins/Hz$^{1/2}$ and a concentration sensitivity $C_{min,i}\cong 3\times10^{18}$ spins/Hz$^{1/2}$m$^3$ could be reached at 4 K. For these oscillators, differently than for OSC1, the frequency noise in the superconducting magnet is significantly larger than the one measured in LN$_2$ and LHe dewar, deteriorating the achievable spin and concentration sensitivity up to two orders of magnitude. The results are summarized in Table \ref{oscillators_results}. 

In conclusion, in this letter we demonstrate the application of low noise microwave LC oscillators based on superconducting resonators to continuous wave ESR spectroscopy. Thanks to the use of superconducting resonators, the oscillators frequency noise is significantly reduced with respect to those based on conducting materials. 

Further frequency noise improvements are possible, as shown by Chaudy et al. \cite{Chaudy2020}, with resonator dimensions and substrate choice (MgO instead of sapphire) possibly playing a role in limiting the resonator intrinsic performance \cite{kermorvant2010}. Nonetheless, a trade off with respect to the final application needs to be considered. Indeed, an additional increase in the resonator dimensions might boost its $Q$-factor, but the resulting increase in the sensitive volume could possibly worsen the spin sensitivity (but improve the concentration sensitivity). 

Due to their relatively low operating frequency (0.6 to 1.7 GHz) and to the set-up induced degradation of the oscillator frequency noise, the noise reduction achieved with the use of superconducting resonators is not yet enough to compensate for the larger ESR signal achievable at higher operating frequencies, resulting in spin sensitivities comparable to the ones obtained with most conducting resonators based oscillators operating in the 8 to 13 GHz range \cite{Yalcin2008,matheoud2018,Schlecker2019}, and lower than the ones operating at frequencies higher than 20 GHz \cite{gualco2014,anders2012,matheoud2017}. However, the studied oscillators show competitive performance in terms of concentration sensitivity with respect to conducting resonators based oscillators operating at frequencies up to 150 GHz. A complete performance comparison with the literature is reported in the in the \textit{supplementary material}. In order to further improve the spin sensitivity, in the future we aim to realize and study superconducting resonators based microwave oscillators operating at 10 GHz and possibly up to 30 GHz, i.e., in a frequency range where superconducting resonators might still allow to obtain significantly higher $Q$-factors with respect to conducting resonators \cite{Blank2017,bonizzoni2016,artzi2022,ghirri2015,dayan2018,zhang2016,asfaw2017,probst2017} and consequently lower noise in the ESR condition. 

Besides their application for ESR spectroscopy, superconducting resonators based microwave oscillators might find applications as local microwave sources in a variety of cryogenic studies, such those related to quantum or classical computing at cryogenic temperatures.

\section*{Supplementary Material}
See the supplementary material for the detailed description of the resonators fabrication and of the oscillators electronics and assembly. Additional measurements and simulations, comparisons with theoretical results and with literature are also provided to support the main findings of this study.

\begin{acknowledgments}
The authors acknowledge fabrication advices from J. Pernollet, J. Dorsaz, A. Toros, N. Piacentini, D. Bouvet and all the staff from the Center of MicroNanoTechnology (CMi) at EPFL in Lausanne, Switzerland, where the fabrication has been carried out. The authors acknowledge N. S. Solmaz, B. Erbas and B. Tandon for fruitful discussions. The financial support from the Swiss National Science Foundation (SNSF) (grant no.: Ambizione PZ00P2\_193361) and the EPFL Center for Quantum Science and Engineering (QSE) (grant no: QSE-CRF2022) are gratefully acknowledged.
\end{acknowledgments}

\section*{Data Availability Statement}
The data that support the findings of this study are available from the corresponding author upon reasonable request.

\section*{Author Declaration}
The authors have no conflicts to disclose.

\section*{Credits}
This article may be downloaded for personal use only. Any other use requires prior permission of the author and AIP Publishing. This article appeared in Appl. Phys. Lett. 126, 154101 (2025) and may be found at \href{https://doi.org/10.1063/5.0260098}{10.1063/5.0260098}.

\bibliography{mybibfile}

\end{document}


\title{Supplementary material Superconducting microwave oscillators as detectors for ESR spectroscopy}

\author{R. Russo}
\email{roberto.russo@epfl.ch}
\affiliation{ 
	Microsystems Laboratory, Ecole Polytechnique Fédérale de Lausanne (EPFL), 1015, Lausanne, Switzerland 
}
\affiliation{ 
	Center for Quantum Science and Engineering, Ecole Polytechnique Fédérale de Lausanne (EPFL), 1015, Lausanne, Switzerland }

\author{A. Chatel}

\affiliation{ 
	Microsystems Laboratory, Ecole Polytechnique Fédérale de Lausanne (EPFL), 1015, Lausanne, Switzerland 
}
\affiliation{ 
	Center for Quantum Science and Engineering, Ecole Polytechnique Fédérale de Lausanne (EPFL), 1015, Lausanne, Switzerland }
\author{N. Brusadin}
\affiliation{ 
	Microsystems Laboratory, Ecole Polytechnique Fédérale de Lausanne (EPFL), 1015, Lausanne, Switzerland 
}
\author{R. Yu}
\affiliation{ 
	Microsystems Laboratory, Ecole Polytechnique Fédérale de Lausanne (EPFL), 1015, Lausanne, Switzerland 
}
\author{R. Farsi}
\affiliation{ 
	Microsystems Laboratory, Ecole Polytechnique Fédérale de Lausanne (EPFL), 1015, Lausanne, Switzerland 
}

\author{H. Furci}
\affiliation{ 
	Microsystems Laboratory, Ecole Polytechnique Fédérale de Lausanne (EPFL), 1015, Lausanne, Switzerland 
}

\author{J. Brugger}
\affiliation{ 
	Microsystems Laboratory, Ecole Polytechnique Fédérale de Lausanne (EPFL), 1015, Lausanne, Switzerland 
}

\author{G. Boero}
\affiliation{ 
	Microsystems Laboratory, Ecole Polytechnique Fédérale de Lausanne (EPFL), 1015, Lausanne, Switzerland 
}
\affiliation{ 
	Center for Quantum Science and Engineering, Ecole Polytechnique Fédérale de Lausanne (EPFL), 1015, Lausanne, Switzerland }
\date{\today}

\maketitle

\section{Process flow for YBCO resonators}									\label{section: YBCO_processes}

\begin{figure*}[ht!]
	\includegraphics[width=\linewidth]{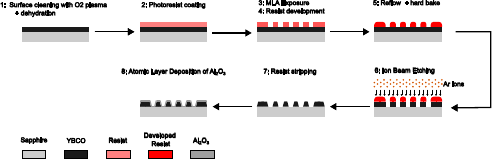} 
	\caption{Fabrication process flow for YBCO standalone resonators. }
	\label{figure: Process_1}
\end{figure*}

The fabrication of the superconducting yttrium barium copper oxide (YBCO) resonators starts with 10$\times$10 mm$^{2}$ chips provided by Ceraco GmbH. These chips consist of a 500 $\mu$m thick r-cut sapphire substrate coated by 300 nm of YBCO on top of a 20 nm buffer layer of CeO$_2$. These films are characterized by a stoichiometric yttrium excess, generating the incorporation of Y$_2$O$_3$ particles, which leads to enhanced pinning and increased $J_c$. Two fabrication processes have been developed, respectively with (see Fig.\ref{figure: Process_2}) and without (see Fig.\ref{figure: Process_1}) metal pads for wire bonding connection to the external feedback electronics.   

\subsection{Process flow for YBCO stand alone resonators}				\label{subsection: YBCO_process_1}

\begin{figure*}[ht!]
	\includegraphics[width=\linewidth]{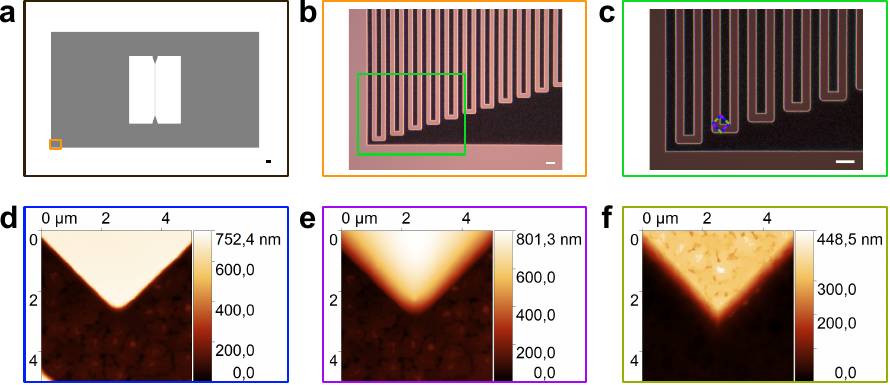} 
	\caption{Fabrication results for standalone resonators. \textbf{a}, Design micrograph of the resonator, scale bar = 100 $\mu$m. \textbf{b, c}, Dark field zoom of the finger region, scale bar = 10 $\mu$m. \textbf{d, e, f}, Atomic force microscope (AFM) scan of the finger region after lithography, reflow and Atomic layer deposition (ALD), respectively.}
	\label{figure: fab_result_composition}
\end{figure*}

The YBCO chips firstly undergo 60 s of O$_2$ plasma cleaning (Tepla 300, 600 W, 400 mL/min O$_2$ flow), then a dehydration step is performed at 160 $^{\circ}$C for 180 s. The chips are then coated with a 700 nm thick AZ 10XT-07 resist (1:1 weight dilution of AZ 10XT-60 from Merck with Propylene glycol methyl ether acetate (PGMEA)). The resist coating is performed using the Sawatec SM-200 manual coater for i-line photoresists (40 s at 6000 rpm). The coating is followed by a soft bake step onto a hot plate (120 s at 120 $^{\circ}$C). The design of the resonator is transferred using maskless direct photolithography (Heidelberg Instruments MLA150, 405 nm wavelength) with an optimized dose of 360 mJ/cm$^2$. Afterwards, the exposed resist is developed in a 1:3.5 diluted AZ 400K KOH based solution for 80 s. After the photolithography steps, the chips are exposed for 10 s to O$_2$ plasma (Tepla 300, 200 W) for a descumming step. Later, in order to prepare the wafer for the etching step, the reflow of the resist is required to avoid fencing during the following Ar milling. The reflow is performed on a hot plate at 125 $^{\circ}$C for 150 s. A hard bake step follows (60 minutes at 85 $^{\circ}$C). The chips are then glued on a support Si dummy wafer (QuickStick 135, Kapton tape) to perform the Ar milling. In order to reduce the heat transferred to the masking resist during the ion beam etching (IBE) and to avoid a temperature rise on YBCO surface, the milling is performed in an intermittent fashion, alternating etching steps and resting/cool-down steps. The final recipe consists of: 4$\times$(8$\times$ (30 s etch + 60 s cool down) + 600 s cool down) + 7$\times$(30 s etch + 60 s cool down)+ 600 s cool down+ 2$\times$(30 s etch + 60 s cool down) + 600 s cool down + 4$\times$(30 s etch + 60 s cool down). The 30 s etching step is performed at an angle of -10$^{\circ}$ and at low plasma power (300 V, 500 mA) using the Veeco Nexus IBE350. The full recipe consists of a total etch time of 22.5 minutes, a total short cool down time (plasma source ON, but etching stopped by metallic shutter) of 45 minutes and a total long cool down time (plasma source OFF) of 60 minutes. After the etching step, the chips are detached from the carrier wafer and the protective resist is removed in a 3 steps subprocess, by alternating dry (O$_2$ plasma, Tepla 300, 600 W, 400 mL/min O$_2$ flow, 60 s), wet (Remover 1165 at 70 $^{\circ}$C for 10 minutes) and again dry processes (O$_2$ plasma, Tepla 300, 600 W, 400 mL/min O$_2$ flow, 60 s). The resist stripping is accomplished following these three steps because: the first removes the "hardened resist crust" formed consequently to the IBE process, the second removes most of the masking resist, and last one removes any possible resist residual. Because of the degradation processes that may occur in presence of water and CO$_2$ infiltrations in the YBCO structure, the  superconducting thin film is protected by a 10 to 50 nm thick layer of  Al$_2$O$_3$ obtained by atomic layer deposition (ALD) at 130 $^{\circ}$C using H$_2$O and trimethylaluminum (TMA) as precursors. This deposition step concludes the process for the fabrication of the standalone YBCO resonators. Fabrication results for YBCO stand alone resonators are reported in Fig.\ref{figure: fab_result_composition}.

\subsection{Process flow for YBCO resonators with connection pads}		\label{subsection: YBCO_process_2}

\begin{figure*}[ht!]
	\includegraphics[width=\linewidth]{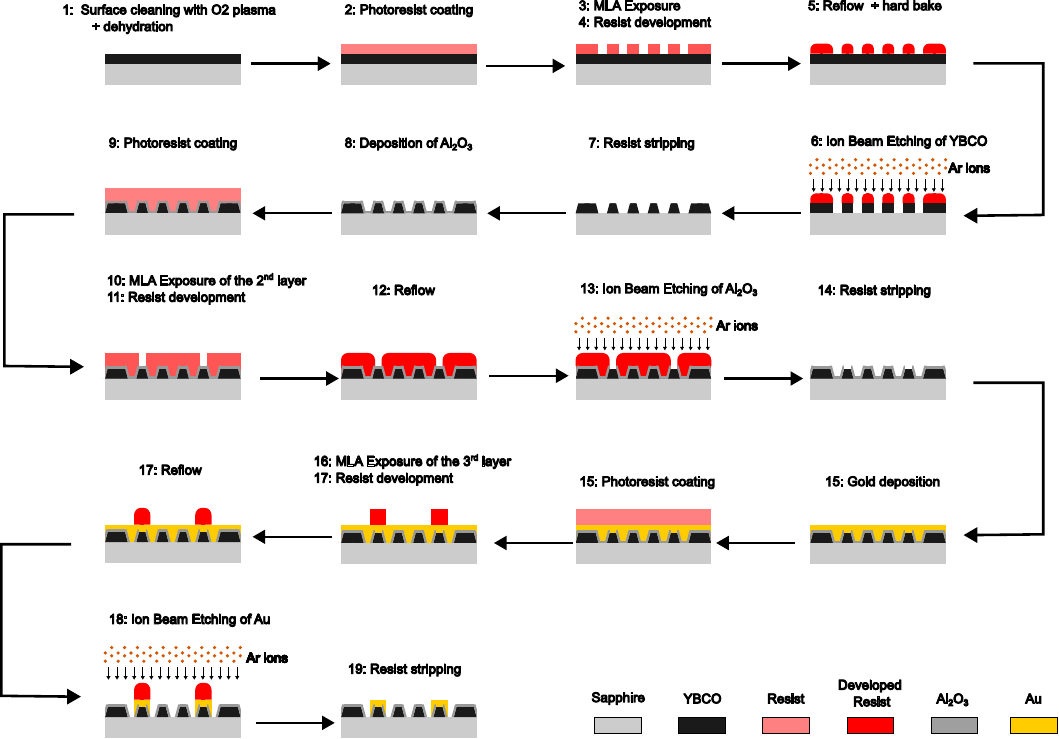} 
	\caption{Fabrication process flow for YBCO resonators with connection pads. }
	\label{figure: Process_2}
\end{figure*}

\begin{figure*}[t!]
	\includegraphics[width=0.95\textwidth]{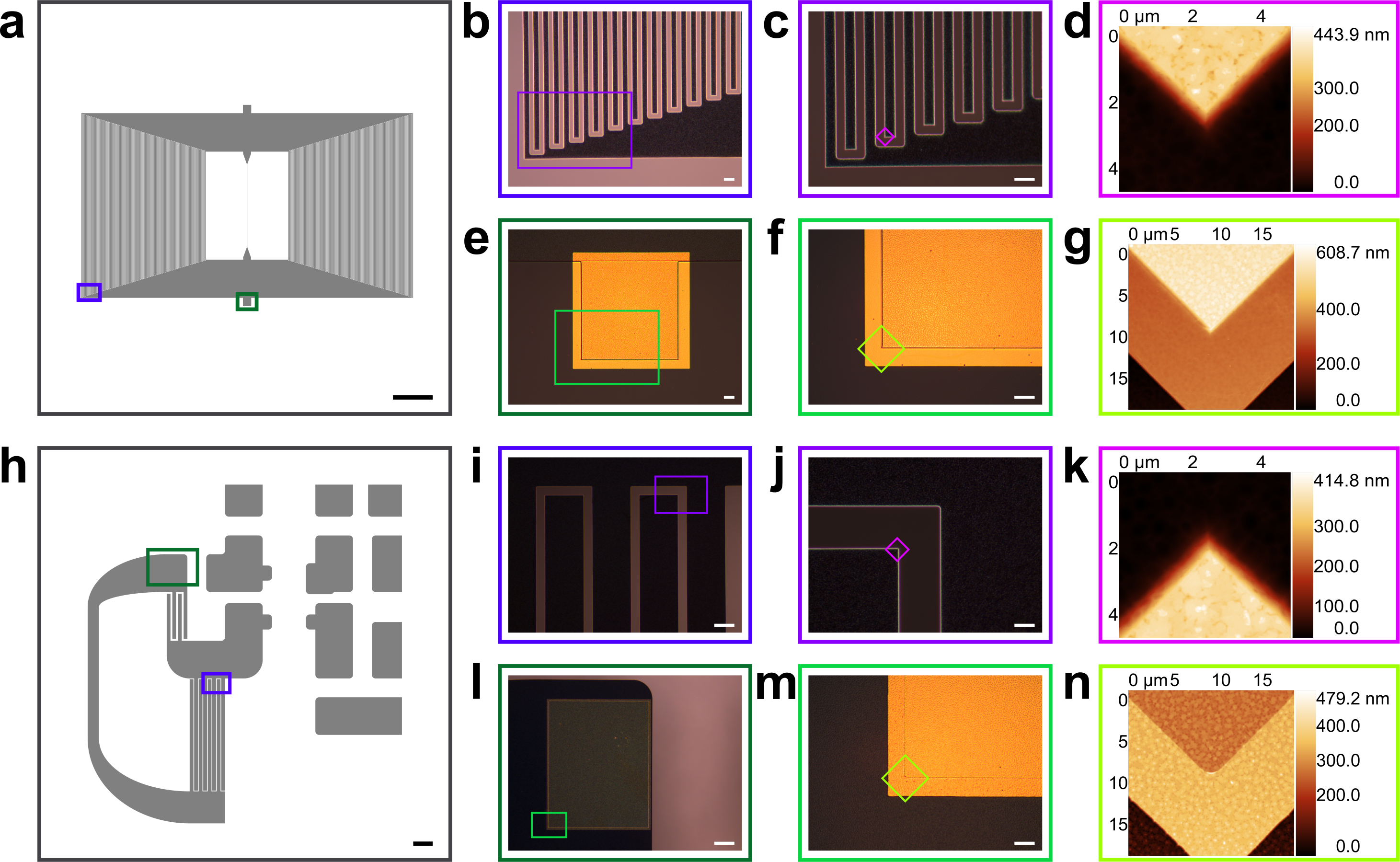}
	\caption{Fabrication results of the YBCO resonators with pads. \textbf{a}, Design of the resonator for OSC1, scale bar = 500 $\mu$m. \textbf{b, c}, Dark field zoom of the finger region, scale bar = 10 $\mu$m. \textbf{d}, Atomic force microscope (AFM) scan of the finger tip after Atomic layer deposition (ALD). \textbf{e, f}, Bright field zoom of the pad region, scale bar = 10 $\mu$m. \textbf{g}, AFM scan of the pad corner at the process end. \textbf{h}, Design of the resonator for OSC2, scale bar = 500 $\mu$m. \textbf{i, j}, Dark field zoom of the finger region, scale bar = 50 $\mu$m and 10 $\mu$m, respectively. \textbf{k}, AFM scan of the finger tip after ALD. \textbf{l}, Dark field zoom of the pad region, scale bar = 100 $\mu$m. \textbf{m}, Bright field zoom of the pad region, scale bar = 10 $\mu$m. \textbf{n}, AFM scan of the pad corner at the process end.}
	\label{figure: fab_result_composition_step_by_step}
\end{figure*}

The second process adds up few steps to the first one to allow for the realization of metallic contact pads on the resonators. These steps are necessary for the wire bonding connection of the resonators with the feedback electronics. More specifically, two photholitography plus etching steps are added to open the Al$_{2}$O$_{3}$ capping layer and to define the metal pads on the resulting aperture. Concerning the second lithography step, the only parameters changed with respect to the first layer are: the used resist, in this case a 2 $\mu$m thick AZ 10XT-20 resist (1:0.42 weight dilution of AZ 10XT-60 from Merck with PGMEA); the coating recipe (40 s $@$ 4000 rpm); the exposure dose (460 mJ/cm$^2$) and the development time (130 s) . All the preparatory procedures before etching are repeated as well, except the hard bake. The etching is again performed with IBE (Veeco Nexus IBE350) in an intermittent fashion. The total time of the recipe changes following the thickness of the Al$_{2}$O$_{3}$ capping layer to etch. More specifically, for a 10 nm thickness the full recipe consists of 5 $\times$ (30 s etch + 60 s cool down), while for a 50 nm thickness the full recipe, which is performed twice to go through the full thickness, consists of 8 $\times$ (30 s etch + 60 s cool down) + 600 s cool down + 3 $\times$ (30 s etch + 60 s cool down). Again the 30 s etching step is performed at an angle of -10$^{\circ}$ and at a low plasma power (300 V, 500 mA), and the cool down steps keep the same parameters as the ones presented in \ref{subsection: YBCO_process_1}. After the resist stripping procedure, performed as in the first layer lithography, Au is sputtered at room temperature (20 $^{\circ}$C) with the Alliance-Concept DP650 single chamber multi target tool. The chamber is kept at a pressure of 5$\times 10^{-3}$ mbar. A DC power of 50 W is applied to the target for 1053 to 1800 s in order to approximately obtain, respectively, a film of about 200 nm and about 340 nm (average deposition rate of 1.9 {\AA}/s). The increase of the Au deposited thickness has improved the ease of the later following bonding process. To define where to leave the Au layer (i.e. the pads) a 3rd layer lithography is performed following the already discussed series of steps. The parameters for this step are: 2 $\mu$m thick AZ 10XT-20 resist by spin coating for 40 s at 4000 rpm, exposure with 420 mJ/cm$^2$ dose and development time of 150 s. Again all the preparatory steps before etching are performed except hard bake. The etching is again performed with IBE (Veeco Nexus IBE350) in an intermittent fashion, with a total recipe time depending on the Au thickness. More specifically, for 200 nm thickness it consists of 7 $\times$ (30 s etch + 60 s cool down), while for 340 nm thickness it consists of 8 $\times$ (30 s etch + 60 s cool down) + 600 s cool down + 3 $\times$ (30 s etch + 60 s cool down). Again the 30 s etching step is performed at an angle of -10$^{\circ}$ and at low plasma power (300 V, 500 mA), and the cool down steps keep the same parameters as the ones presented in \ref{subsection: YBCO_process_1}. The resist stripping procedure, performed as before, concludes the process. Fabrication results for YBCO resonators with pads for connections are reported in Fig.\ref{figure: fab_result_composition_step_by_step}.

\section{Process flow for NbTi resonators}							\label{section: NbTi_process_1}

\begin{figure*}[ht!]
	\includegraphics[width=\linewidth]{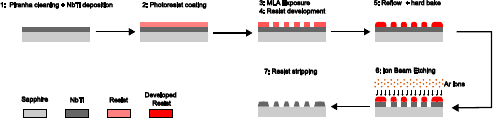} 
	\caption{Fabrication process flow for NbTi resonators. }
	\label{figure: Process_nbti}
\end{figure*}

The fabrication of the superconducting NbTi resonators starts with 4 inch sapphire wafers provided by Siegert Wafer GmbH. The sapphire substrate is firstly cleaned in piranha solution (H$_2$SO$_4$ 96$\%$ activated by H$_2$O$_2$ 30$\%$) at 100 $^{\circ}$C for 10 minutes. The Nb$_{0.5}$Ti$_{0.5}$ layer is DC sputtered at room temperature (20 $^{\circ}$C) with the Alliance-Concept Dp650 single chamber multi target tool. The chamber is kept at a pressure of 5$\times 10^{-3}$ mbar. A DC power of 250 W is applied to the target for 455 s in order to obtain a film of approximately 150 nm (average deposition rate of 3.3 {\AA}/s). After NbTi sputtering , the wafer is coated with a 750 nm thick AZ 10XT-07 resist (1:1 weight dilution of AZ 10XT-60 from Merck with PGMEA). The coating is performed using the Suss MicroTec ACS200 Gen3 cluster for i-line photoresists. The coating recipe is optimized in order to obtain a good uniformity of the polymer \cite{Russo2023}. The coating recipe consists of the following steps: dehydration for 180 s at 160 $^{\circ}$C, cool down for 90 s at 22 $^{\circ}$C, coating for 72 s up to 5650 rpm, soft bake for 120 s at 120 $^{\circ}$C and cool down for 15 s at 22 $^{\circ}$C. The design of the resonator is transferred using maskless direct photolithography (Heidelberg Instruments MLA150, 405 nm wavelength) with an optimized dose of 160 mJ/cm$^2$. Afterwards, the exposed resist is developed in a 1:3.5 diluted AZ 400K KOH based solution. After the photolithography steps, the wafer is exposed for 10 s to O$_2$ plasma (Tepla GiGAbatch, 200 W, 200 sccm O$_2$ flow, 0.5 mbar pressure) for a descumming step. Later, in order to prepare the wafer for the etching step, the reflow of the resist is required to avoid fencing during the following Ar milling. The reflow is performed on a hot plate at 135 $^{\circ}$C for 120 s. In order to reduce the heat transferred to the masking resist during the ion beam etching (IBE), the milling is performed in an intermittent fashion, alternating etching steps and resting/cool-down steps. The final recipe consists of 8 $\times$ (30 s etch + 60 s cool down) + 600 s cool down + 8 $\times$ (30 s etch + 60 s cool down). The 30 s etching step is performed at an angle of -10$^{\circ}$ and at low plasma power (300 V, 500 mA) using the Veeco Nexus IBE350. Concluded the etching step, the resist is removed by alternating dry (O$_2$ plasma, Tepla GiGAbatch, 600 W, 400 sccm O$_2$ flow, 0.8 mbar pressure, 1 minute) and wet processes (Remover 1165 at 70 $^{\circ}$ C for 10 minutes). The finalized wafer is then protected with thick resist (Merck AZ 40XT, 20 $\mu$m) to perform the mechanical dicing into chips (Disco DAD321). A resist as thick as 20 $\mu$m has been chosen since trials at lower thicknesses ($\sim$ 4-5 $\mu$m) have shown small damages to the material surface. The protective resist is removed after dicing in acetone and the samples rinsed in IPA and water. 

\section{Assembly of the oscillators}									\label{section: assembly_osc}

We have fabricated and characterized three different oscillators based on superconducting resonators: oscillator 1 (OSC1), oscillator 2 (OSC2), and oscillator 3 (OSC3). In the following, the assembly of the three oscillators is reported in details.  

\subsection{OSC1}									\label{subsection: ass_osc_01}

A printed circuit board (PCB) is mechanically modified to host the 10$\times$10 mm$^2$ chip on which the resonator is fabricated. This modification consists of performing a squared aperture of approximately 6$\times$6 mm$^2$ on the top face, and to enlarge this aperture to approximately 11$\times$11 mm$^2$ on the bottom. Once the aperture is ready, first the needed surface mounting device (SMD) components are soldered on the PCB, then the chip is glued in position with several step of glue positioning (non conductive glue EPO-TEK H70E) and baking (80 $^{\circ}$C for 90 minutes). A small piece of PCB is positioned on the back of the chip with the same glue to avoid possible loss of the chip due to thermal shock cracking of the glue. Finally, a sealing layer of glue (conductive glue EPO-TEK H20E-FC) is used, far from the resonator, on the borders of the PCBs, to keep in place the additional piece of PCB. This last glue requires an additional baking step (80 $^{\circ}$C for 45 minutes). After the conclusion of the physical assembly of the components and chip onto the PCB, it is required to bond the resonator pads to the corresponding connections on PCB. The bonding is performed with Al wires of 33 $\mu$m diameter by using the TPT HB10 wedge and ball bonder tool.

\subsection{OSC2 and OSC3}									\label{subsection: ass_osc_02_osc_03}

In the case of OSC2 and OSC3, which are based on a pHEMT transistor, the assembly approach is different with respect to \ref{subsection: ass_osc_01}. Indeed, the PCB does not require any mechanical modification. As first step the fabricated chip is glued on top of the PCB with a thin layer of non conductive glue (EPO-TEK H70E), which avoid the creation of a GND plane on the back of the chip. Again, to be sure not to lose the chip because of mechanically instabilities of the non conductive glue as a consequence of thermal shock, a small amount of conductive glue (EPO-TEK H20E-FC) is used at three out of the four corners of the chip. Because of the limited space on the chip, and to avoid crushing any SMD components during the bonding procedure, the bonding wires are placed before the electronic SMD components. Also, in this case, several bonding wires are connected in parallel to reduce the inductance of the connection. More specifically, 2 to 5 Al bonding wires of 25 $\mu$m diameter are placed by using the F\&S BONDTEC 5630i automatic wire bonder tool. After the placing of the bondings, the passive and active SMD components are glued in position using a tiny amount of conductive glue. 

\section{Details on the circuit components}									\label{section: ele_components}

\begin{figure*}[h!] 
	\includegraphics[width=\linewidth]{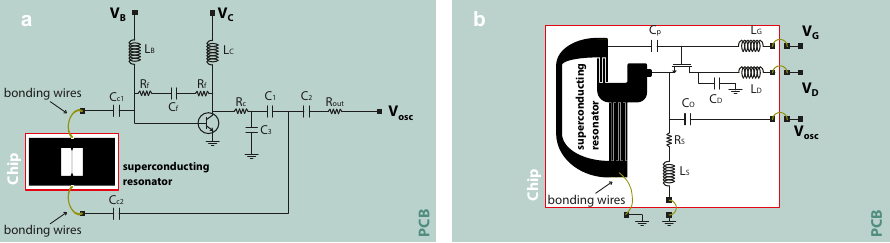}
	\caption{Electronic circuits of the oscillators. \textbf{a}, Circuit for HBT based OSC1. \textbf{b}, Circuit for pHEMT based OSC2 and OSC3.}
	\label{figure: circuit_bjt_and_hemt}
	\makeatletter
	\let\save@currentlabel\@currentlabel
	\edef\@currentlabel{\save@currentlabel a}\label{figure: circuit_bjt_and_hemt:a}
	\edef\@currentlabel{\save@currentlabel b}\label{figure: circuit_bjt_and_hemt:b}
	\makeatother
\end{figure*}

\begin{center}
	\begin{table}[!h]
		\begin{tabular}{cccccccccccc}
			\thead{\textit{Device}}						& \thead{\textit{L$_{B}$}  	\text{[nH]}} 	& \thead{\textit{L$_{C}$}  	\text{[nH]}} 	& \thead{\textit{R$_{f}$}  	\text{[K$\Omega$}]}		& \thead{\textit{C$_{f}$}  	\text{[pF]}} 	& \thead{\textit{R$_{c}$}  	\text{[$\Omega$]}} 		& \thead{\textit{C$_{1}$} 	\text{[pF]}} 	& \thead{\textit{C$_{2}$}  	\text{[pF]}} 	& \thead{\textit{C$_{3}$}  	\text{[pF]}} 	& \thead{\textit{C$_{c1}$}  	\text{[pF]}}	& \thead{\textit{C$_{c2}$}  	\text{[pF]}} 	& \thead{\textit{R$_{out}$}  	\text{[$\Omega$]}} 	\\	\hline \hline
			\thead{\textit{OSC1}} 					& \thead{56} 	& \thead{390} 	& \thead{1} 	& \thead{100 } 	& \thead{35} 	& \thead{100} 	& \thead{100} 	& \thead{18} 	& \thead{4} 		& \thead{4} 		& \thead{500}  \\
		\end{tabular}
		\caption{Value of the components for the HBT based oscillator circuit (OSC1).}
		\label{table: components_bjt}
	\end{table}	
\end{center}

\begin{center}
	\begin{table}[!h]
		\begin{tabular}{ccccccc}
			\thead{\textit{Device}}		& \thead{\textit{L$_{G}$}  	\text{[nH]}} 	& \thead{\textit{L$_{S}$}  	\text{[nH]}} 	& \thead{\textit{R$_{S}$}  	\text{[$\Omega$}]}		& \thead{\textit{C$_{p}$}  	\text{[pF]}} 	& \thead{\textit{C$_{o}$} 	\text{[pF]}} 	& \thead{\textit{C$_{D}$}  	\text{[pF]}}  	\\	\hline \hline
			\thead{\textit{OSC2}} 	& \thead{56} 	& \thead{56} 	& \thead{5} 	& \thead{33} 	& \thead{100} 	& \thead{100} 	  \\
			\thead{\textit{OSC3}} 	& \thead{56} 	& \thead{56} 	& \thead{5} 	& \thead{33} 	& \thead{100} 	& \thead{100} 	  \\
		\end{tabular}
		\caption{Value of the components for the pHEMT based oscillator circuits (OSC2 and OSC3).}
		\label{table: components_hemt}
	\end{table}	
\end{center}

As previously mentioned, two different electronics configuration are used, the former referred as "blocks configuration" based on an HBT active device (Infineon BFP650) and the latter, a more conventional Colpitts configuration, based on a pHEMT (Skyworks SKY65050-372LF) active device. The electronic components chosen for the first configuration (OSC1) are reported in Table \ref{table: components_bjt} and the corresponding circuit is shown in Fig.\ref{figure: circuit_bjt_and_hemt:a}. The electronic components chosen for the second configuration (OSC2 and OSC3) are reported in Table \ref{table: components_hemt} and the corresponding circuit is shown in Fig.\ref{figure: circuit_bjt_and_hemt:b}.

\section{Frequency noise comparison: superconductors vs normal conductors based oscillators}
\label{section: comparison_noise_materials}

\begin{figure*}[ht!]
	\centering 
	\includegraphics[width=\linewidth]{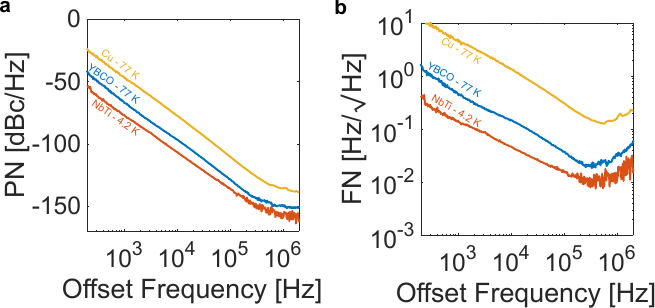}
	\caption{Comparison of the noise performance of the pHEMT based oscillator with respect to the resonator material. The comparison refers to OSC2 (pHEMT YBCO), OSC3 (pHEMT NbTi), and an equivalent oscillator with the same resonator design realized with Cu on PCB (OSC Cu). \textbf{a}, Phase noise comparison. \textbf{b}, Frequency noise comparison.}
	\label{figure: hemt_mat_comp}
\end{figure*}

To demonstrate the improvements in frequency noise when realizing the oscillators using superconducting resonators instead of conducting resonator we compared the best achieved noise of OSC2 (YBCO pHEMT) and OSC3 (NbTi pHEMT) with an equivalent oscillator (OSC Cu) with the resonator, having the same shape and design, but realized with Cu (35 $\mu$m thick) on a PCB. OSC2 is characterized in a dewar of liquid nitrogen (77 K) as well as OSC Cu, while OSC3, being realized in NbTi, is characterized in a dewar of liquid helium (4.2 K). As shown in Fig.\ref{figure: hemt_mat_comp}, the oscillator based on a superconducting resonator achieves a frequency noise which is one order of magnitude better than the one achieved with the conducting resonator.

\section{ESR measurements Setup}
\label{section: esr_setup}

\begin{figure*}[ht!] 
	\includegraphics[width=0.9\linewidth]{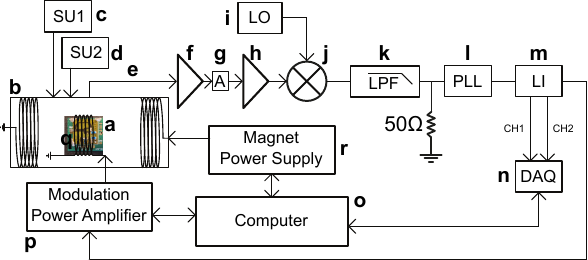} 
	\caption{Block diagram of the experimental set-up for ESR measurements of the oscillators in the superconducting magnet. \textbf{a}, Oscillator to test. \textbf{b}, Superconducting magnet with a variable temperature insert, 1.4 to 300 K, 0 to 9.4 T (Cryogenic Ltd). \textbf{c}, \textbf{d}, Source-meter units for oscillator bias (Keithley 2400). \textbf{e}, RF cables (Huber+Suhner, Sucoflex 104). \textbf{f}, RF Amplifier (Mini-Circuits ZX60-V83-S+ or Mini-Circuits ZLN-1000 LN+). \textbf{g}, Attenuator 4dB. \textbf{h}, RF Amplifier (Mini-Circuits ZLN-1000 LN+, ). \textbf{i}, Signal generator for downconversion (Marconi Instruments 2030, SRS SG384, or HP 8657B). \textbf{j}, RF mixer (Mini-Circuits ZFM-2000+ or Mini-Circuits ZEM-4300+). \textbf{k}, Low-pass filter (Mini-Circuits BLP-10.7+). \textbf{l}, Phase-Locked Loop, custom design. \textbf{m}, Lock-In Amplifier (EG\&G 7260). \textbf{n}, Data acquisition board (NI, PCIe-6353). \textbf{o}, PC with NI LabView software. \textbf{p}, Power amplifier (Rohrer, PA508X, ). \textbf{q}, Modulation coil, custom design. \textbf{r}, Superconducting magnet power supply (Cryogenic Ltd).}
	\label{figure: meas_setup}
\end{figure*}

The measurement chain to acquire the ESR spectra unfolds as following. The oscillator is inserted in the variable temperature insert of a superconducting magnet. The oscillator is biased through the use of two source-meter units (Keithley 2400), and its output signal routed out and amplified. The amplified signal is downconverted to about 10 MHz using a frequency mixer. The 10 MHz signal is delivered to a custom phase-locked-loop (PLL) which is used as frequency-to-voltage converter. The PLL has a noise limit of about 10 mHz/Hz$^{1/2}$. The signal at the output of the PLL is routed to a lock-in amplifier which performs the synchronous demodulation of the input signal at the frequency of the magnetic field modulation applied with a custom modulation coil. The current set-up allows also to change the $\phi$ angle (i.e., the angle in the xy plane, where z is the direction of the static magnetic field), but not the $\theta$ angle. The $\phi$ angle can be changed by rotating the probe together with the sample, this means that the direction of B$_1$ (which is in the xy plane) with respect to the sample is fixed.

\section{Frequency noise degradation inside the superconducting magnet} \label{section: noise_chain}

\begin{figure*}[ht!]
	\centering 
	\includegraphics[width=\linewidth]{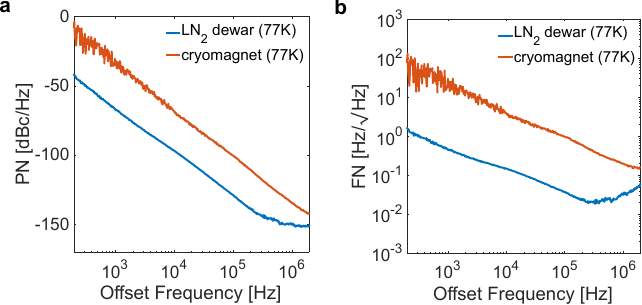} 
	\caption{Influence of the measurement environment. \textbf{a}, Phase noise comparison for OSC2 in a dewar of liquid nitrogen (77 K) and in the cryomagnet (77 K). \textbf{b}, Frequency noise comparison for OSC2 in a dewar of liquid nitrogen (77 K) and in the cryomagnet (77 K). All measurements taken at the same bias point, with V$_{G}$ =-0.661 V and I$_{D}$ = 5 mA.}
	\label{figure: noise_in_out_magnet}
\end{figure*}

As mentioned in the main text, we experimentally observed that the oscillators frequency noise spectral density FN is strongly influenced by the experimental environment in which the oscillator frequency noise is measured. We performed experiments with the oscillators immersed in dewars of liquid nitrogen (77 K) and liquid helium (4.2 K) as well as with the oscillators in contact with static He gas in the variable temperature insert of a superconducting magnet (1.4 to 325 K, 0 to 9.4 T, Cryogenic Ltd.). The sensitivity to the measurement environment changes for the different oscillators, with OSC1 being the less influenced in terms of noise increase. The difference in frequency noise spectral density between the experimental conditions can be as large as 40 dBc (i.e., a factor of 100) as it can be seen in Fig.\ref{figure: noise_in_out_magnet} for OSC2. This large increase in the frequency noise in the superconducting magnet is not caused by the mechanical vibrations produced by the cooling system based on a Gifford McMahon (GM) coldhead. We performed experiments with the cryocooler temporarily switched off, observing only a reduction of the noise at frequency below 100 Hz, in particular at the main mechanical vibration frequencies of 1 Hz and 2 Hz, and no significant changes at higher frequencies, including those used for field modulation in the ESR experiments. We observed also no changes by switching off the pumps used to circulate the He gas as well as the power supplies of all superconducting coils (main coil, sweep coil, shims coils) and resistive coils (modulation coil). The experiments shown in Fig.\ref{figure: noise_in_out_magnet} has been performed in both cases at approximately zero magnetic field. In the dewar environment, the magnetic field is that produced by the Earth. In the superconducting magnet, a residual magnetic field of maximum few mT cannot be excluded. However, as reported in Fig.2 of the main text, the frequency noise does not depend on the magnetic field up to, at least 7 T when working at 3 K. This allows to reasonably exclude that the very significant increase in the frequency noise inside the superconducting magnet, recorded at all tested temperatures for ESR, from 70 K down to 4 K, and here presented as comparison in Fig.\ref{figure: noise_in_out_magnet}, is due to the residual magnetic field inside the superconducting magnet. A clear difference between the two experimental environments is the cooling medium in contact with the oscillator: liquid nitrogen or liquid helium in the dewars, static helium gas in the superconducting magnet. At 77 K, the thermal conductivity of liquid nitrogen is about 0.16 W/mK \cite{Gopal2023} whereas the one of helium gas is 0.05 W/mK \cite{Gopal2023}. At 4.2 K and atmospheric pressure, the thermal conductivity of liquid helium is about 0.02 W/mK \cite{linkHe} whereas the one of helium gas is 0.01 W/mK \cite{Gopal2023}. Another difference between the two experimental conditions is the temperature profile along the probe. However, at the moment, we have no clear indications that these differences are responsible for the increased frequency noise in the superconducting magnet.

\section{ESR signal amplitude}		\label{section: ESR_amplitude_theory}

In the following, we give an approximated expression for the ESR signal amplitude as variation of the oscillation frequency of an oscillator coupled with an ensemble of electron spins. From the steady-state solution of the Bloch equation, the susceptibility of an ESR sample having a single homogeneously broadened line placed in a static magnetic field ${B_0}{\bf{\hat z}}$ and a microwave
magnetic field $2{B_1}\cos \left( {\omega t} \right){\bf{\hat x}}$ can be written as \cite{Yalcin2008, matheoud2018}
\begin{equation}
\chi ' =  - \frac{1}{2}\frac{{\left( {\omega  - {\omega _0}} \right)T_2^2}}{{1 + T_2^2{{\left( {\omega  - {\omega _0}} \right)}^2} + {\gamma ^2}B_1^2{T_1}{T_2}}}{\omega _0}{\chi _0}
\end{equation}

\begin{equation}
\chi '' = \frac{1}{2}\frac{{{T_2}}}{{1 + T_2^2{{\left( {\omega  - {\omega _0}} \right)}^2} + {\gamma_e ^2}B_1^2{T_1}{T_2}}}{\omega _0}{\chi _0}
\end{equation}
where $\chi  = \chi ' - j\chi ''$  is the sample complex susceptibility, $\omega_0=\gamma_e B_0$ is the Larmor frequency , $\gamma_e$ is the electron gyromagnetic ratio, $B_1$ is the microwave magnetic field, $T_1$ and $T_2$ are the longitudinal and transversal relaxation times of the sample under investigation, and $\chi_0$ is the static magnetic susceptibility. The static magnetic susceptibility, in the Curie's law approximation, can be written as ${\chi _0} = {\mu _0}N{\mu ^2}/3kT$ , where $N$ is the spin density (in m$^{-3}$), $T$ is the sample temperature (in K), $\mu$ is the total magnetic moment, $\mu_0$ is the vacuum permeability,  and $k$ is the Boltzmann constant.  For a spin system with g-factor $g\cong2$ and spin quantum number $S = 1/2$, the electron gyromagnetic ratio is $\gamma_e\cong 2\pi\times 28$ GHz/T, and the static magnetic susceptibility is $\chi_0$ = $\mu_0N\gamma_e^2\hbar^2 / 4 kT$.  The impedance of a coil filled with a material having susceptibility $\chi$ is
\begin{equation}
{Z_\chi } = j\omega L(1 + \eta \chi ) + R = j\omega {L_\chi } + {R_\chi }
\end{equation}
where ${L_\chi } = L + L\eta \chi '$ and  ${R_\chi } = R + \omega L\eta \chi ''$. $R$ and $L$ are the inductance and the resistance of the coil without the sample, respectively. For $\omega  \cong {\omega _0}$, the presence of the sample modifies the reactive as well as the resistive part of the coil impedance. The filling factor  $\eta$ is given by
\begin{equation}
\eta  = \frac{{\int\limits_{{V_S}} {{{\left| {{B_{u,xy}}} \right|}^2}dV} }}{{\int\limits_V {{{\left| {{B_u}} \right|}^2}dV} }} = \frac{{\int\limits_{{V_S}} {{{\left| {{B_{u,xy}}} \right|}^2}dV} }}{{{\mu _0}L}}
\label{eq: eta}
\end{equation}
where $B_{u,xy}$ is the field produced by a unitary current in the resonator along the direction perpendicular to the static magnetic field $B_0$, $V_s$ is the sample volume, $V$ the entire space where $B_u$ is not zero.
The oscillation frequency of an oscillator inductively coupled with an ensemble of electron spins, can be written as
\begin{equation}
{\omega _{LC\chi }} \cong \frac{1}{{\sqrt {{L_\chi }C} }} \cong \frac{{{\omega _{LC}}}}{{\sqrt {1 + \eta \chi '} }}
\end{equation}
where ${\omega _{LC}} \cong 1/\sqrt {LC}$ is the unperturbed oscillation frequency and  $C$ is the effective capacitance. In condition of negligible saturation (i.e., for ${\gamma_e ^2}B_1^2{T_1}{T_2} << 1$ ) and for $\eta \chi ' << 1$ , the variation of the oscillator frequency is given by \cite{Yalcin2008, matheoud2018}
\begin{equation}
\Delta {\omega _{LC}} \cong \frac{1}{2}{\omega _{LC}}\eta \chi '
\end{equation}
Hence, from Eq. 1, the peak-to-peak variation of the oscillator frequency is
\begin{equation}
\Delta {\omega _{LC,{\rm{pp}}}} \cong \frac{1}{4}\omega _{LC}^2{T_2}\eta {\chi _0}
\end{equation}
or, expressed in Hz, is
\begin{equation}
\Delta {f_{LC,{\rm{pp}}}} \cong \frac{1}{{2\pi }}\Delta {\omega _{LC,{\rm{pp}}}} =  \frac{\pi }{2}f_{LC}^2{T_2}\eta {\chi _0} 
\label{eq: delta_f_pp}
\end{equation}

where f$_{LC}$ is the unperturbed oscillation frequency of the oscillator (in Hz).\cite{Yalcin2008,anders2012,gualco2014,matheoud2017,matheoud2018}

\begin{center}
	\begin{table}[!h]
		\begin{tabular}{ccccccc}
			\thead{\text{Device}}		& \thead{\text{$T$}  	\text{[K]}} & \thead{\text{$f_{LC}$}  	\text{[GHz]}} 	& \thead{\text{$L$}  	\text{[nH]}} 	& \thead{\text{$\eta$}  	\text{[-]}}		& \thead{\text{$\Delta {f_{LC,{\rm{pp},meas}}}$}  	\text{[kHz]}} 	& \thead{\text{$\Delta {f_{LC,{\rm{pp},calc}}}$} 	\text{[kHz]}} 	 	\\	\hline \hline
			\thead{\text{OSC1}} 	& \thead{\text{4.3} \\ \text{10} \\ \text{30} \\ \text{70} } 	& \thead{$\sim$0.6}  & \thead{1.1}	&\thead{\text{3.4$\times10^{-4}$}}  	& \thead{\text{25*} \\ \text{25} \\ \text{8.7} \\ \text{4.3} }  	& \thead{\text{61} \\ \text{26} \\ \text{8.8} \\ \text{3.7}} 		  \\ \hline
			\thead{\text{OSC2}} 	& \thead{\thead{\text{4} \\ \text{10} \\ \text{30} \\ \text{70} }} 	& \thead{$\sim$1.7} & \thead{6} 	& \thead{2.3$\times10^{-5}$} 	& \thead{\text{6.9*} \\ \text{8.1} \\ \text{3.5} \\ \text{1.6} }  	& \thead{\text{33} \\ \text{13} \\ \text{4.4} \\ \text{1.9} } 		  \\ \hline
			\thead{\text{OSC3}} 	& \thead{\text{4}} 	& \thead{$\sim$1.7}  & \thead{6} 	& \thead{8.7$\times10^{-5}$}	& \thead{28*} 	& \thead{115} 		  \\ \hline
		\end{tabular}
		\caption{Comparison between the measured and the computed (Eq.(\ref{eq: delta_f_pp}) ESR signal amplitudes. The measured ESR signal amplitude obtained by field modulation and synchronous demodulation (i.e., lock-in detection) is scaled considering the field modulation amplitude with respect to the linewidth. For a field modulation amplitude of 0.4 G and a linewidth of 1 G, as in our experimental conditions, the conversion factor is 0.3 (i.e., the measured signal at the output of the lock-in is divided by 0.3). The estimation of the conversion factor is performed numerically. The values indicated with an asterisk (*) refers to measured values outside of the range of validity of Curie's law (Eq.(\ref{eq: delta_f_pp}) assumes that the Curie's law is valid). The values of the transversal relaxation time $T_2$ and the spin density $N$ in the computation of the signal amplitude are $T_2=$ 100 ns and $N$ = 1.5$\times$10$^{27}$ m$^{-3}$, as reported for BDPA \cite{azuma1994}. $B_{u,xy}$ in Eq.(\ref{eq: eta}) is approximated with the one produced by a infinitely long wire (for OSC1) and the one generated by a rectangular loop (for OSC2 and OSC3).}
		\label{table: ESR_amp_comparison}
	\end{table}	
\end{center}

As shown in in Table \ref*{table: ESR_amp_comparison}, all measured spectra show a frequency variation within a factor of two with respect to the theoretical value obtained with this equation, except at 4 K where the experimentally measured ESR signal amplitude does not follow anymore the Curie's law. The values of $L$ are calculated using Rosa's formula \cite{Rosa1908} and including the kinetic inductance contribution, even though negligible, as $L_K(T) = \mu_0 \lambda^2(T)(l/wd)$, where $\lambda$(T) is the penetration depth and $l$, $w$ and $d$ are respectively length, width and thickness of the superconducting material \cite{London1935}.

\section{Definition of spin sensitivity and concentration sensitivity}		\label{section: ESR_spin_sensitivity}

In the main text are reported the performance of the three realized oscillators in terms of spin sensitivity and concentration sensitivity. In this section, the equations used to estimate these quantities are detailed. The spin sensitivity (in spins/Hz$^{1/2}$) is defined as $N_{min} = 3N_s/SNR$, where $N_s$ is the number of spins in the sample and $SNR$ is the signal-to-noise ratio (in Hz$^{-1/2}$). The number of spin in the sample is  $N_s=\rho_s V_S$, where $\rho_s$ is the spin density (in spins/m$^{3}$) and $V_S$ is the sample volume (in m$^3$). The signal-to-noise ratio is defined as $SNR = \Delta {f_{LC,{\rm{pp}}}}/FN$,  where $\Delta {f_{LC,{\rm{pp}}}}$ is the peak-to-peak frequency variation of the oscillator frequency caused by the ESR phenomenon (in Hz, see previous section) and $FN$ is the frequency noise spectral density of the oscillator (in Hz/Hz$^{1/2}$).

\section{Performance comparison with literature}		\label{section: performance_comp}

\begin{table*}[!ht]
	\begin{tabular}{rrrrrrrrrrrr}

		\thead{\text{Device} \\ \text{reference}} & \thead{\text{Frequency} \\ \text{[GHz]}} &  \thead{\text{Temperature} \\ \text{[K]}} & \thead{\text{OF} \\ \text{[kHz]}} &  \thead{\text{FN @ OF} \\ \text{[mHz/Hz$^{1/2}$]}} &  \thead{\text{N$_{min}$} \\ \text{[spins/Hz$^{1/2}$]}} & \thead{\text{Approx.}  \text{effective} \\ \text{volume} \text{[mm$^{3}$]}}  & \thead{\text{C$_{min}$} \\ \text{[spins/Hz$^{1/2}$m$^3$]}}  \\ \hline \hline
		
		\thead{\text{OSC 01 (this work)}} & \thead{\text{0.6}} & \thead{\text{4.3}} & \thead{\text{25}} & \thead{\text{15}} & \thead{ \text{1$\times$10$^{10}$}} & \thead{\text{0.7}} & \thead{ \text{2$\times$10$^{19}$}} \\ \hline
		
		\thead{\text{OSC 02 (this work)}} & \thead{\text{1.7}}   & \thead{\text{77}} & \thead{\text{249}} & \thead{\text{19}} & \thead{ \text{5$\times$10$^{11}$}} & \thead{\text{19}}& \thead{ \text{3$\times$10$^{19}$}} \\ \hline
		
		\thead{\text{OSC 03 (this work)}} & \thead{\text{1.7}} & \thead{\text{4.2}} & \thead{\text{77}} & \thead{\text{9}} & \thead{ \text{5$\times$10$^{10}$}} & \thead{\text{19}} & \thead{ \text{3$\times$10$^{18}$}} \\ \hline
		
		\thead{\text{ \cite{Yalcin2008}}} & \thead{\text{8.4} \\\text{9.4}}  & \thead{\text{300}} & \thead{\text{20}} & \thead{\text{30000}} & \thead{ \text{3$\times$10$^{10}$}} & \thead{\text{3$\times$10$^{-3}$}} & \thead{ \text{1$\times$10$^{22}$}} \\ \hline
		
		\thead{\text{ \cite{matheoud2018}}} & \thead{\text{11.2}}    & \thead{\text{300} \\ \text{10}} & \thead{\text{100}} & \thead{\text{17000} \\ \text{4000}} & \thead{ \text{8$\times$10$^{10}$}\\ \text{2$\times$10$^{9}$}} & \thead{\text{3.4$\times$10$^{-2}$}} & \thead{ \text{2$\times$10$^{21}$}\\  \text{6$\times$10$^{19}$}} \\ \hline
		
		\thead{\text{\cite{matheoud2018}}} & \thead{\text{10.1}}    & \thead{\text{300} \\  \text{10}} & \thead{\text{150}} & \thead{\text{2000} \\ \text{4000}} & \thead{ \text{4$\times$10$^{8}$}\\ \text{3$\times$10$^{7}$}} & \thead{\text{4.1$\times$10$^{-3}$}} & \thead{ \text{1$\times$10$^{20}$}\\ \text{7$\times$10$^{18}$}} \\ \hline
		
		\thead{\text{\cite{Schlecker2019} }} & \thead{\text{12.7}}    & \thead{\text{300}} & \thead{\text{100}} & \thead{\text{1000}} & \thead{ \text{3$\times$10$^{9}$}} & \thead{\text{8$\times$10$^{-3}$}} & \thead{ \text{4$\times$10$^{20}$}} \\ \hline
		
		\thead{\text{\cite{gualco2014}}} & \thead{\text{20}}  & \thead{\text{300}\\ \text{4}} & \thead{\text{100}\\ \text{10}} & \thead{\text{20000}\\ \text{1000}} & \thead{ \text{1$\times$10$^{8}$}\\  \text{1$\times$10$^{6}$}} & \thead{\text{1.1$\times$10$^{-3}$}} & \thead{ \text{9$\times$10$^{19}$}\\  \text{9$\times$10$^{17}$}} \\ \hline
		
		\thead{\text{\cite{anders2012}}} & \thead{\text{27}}  & \thead{\text{300}} & \thead{\text{100}} & \thead{\text{20000}} & \thead{ \text{2$\times$10$^{8}$}} & \thead{\text{8$\times$10$^{-4}$}} & \thead{ \text{3$\times$10$^{20}$}} \\ \hline
		
		\thead{\text{\cite{matheoud2017}}} & \thead{\text{50}}   & \thead{\text{300}} & \thead{\text{100}} & \thead{\text{90000}} & \thead{ \text{1$\times$10$^{8}$}} & \thead{\text{6.8$\times$10$^{-4}$}} & \thead{ \text{2$\times$10$^{20}$}} \\ \hline
		
		\thead{\text{\cite{matheoud2017}}} & \thead{\text{92}}    & \thead{\text{300}} & \thead{\text{100}} & \thead{\text{300000}} & \thead{ \text{4$\times$10$^{7}$}} & \thead{\text{1.4$\times$10$^{-4}$}} & \thead{ \text{3$\times$10$^{20}$}} \\ \hline
		
		\thead{\text{\cite{matheoud2017}}} & \thead{\text{146}}   & \thead{\text{300}} & \thead{\text{100}} & \thead{\text{700000}} & \thead{ \text{2$\times$10$^{7}$}} & \thead{\text{3.6$\times$10$^{-5}$}} & \thead{ \text{6$\times$10$^{20}$}} \\ \hline 
	\end{tabular}
	\caption{Comparison of the performance of the oscillators studied in this work with respect to microwave oscillators based on normal metal resonators previously reported in literature. For each device are reported: the reference, the operating frequency, the operating temperature, the frequency noise, the spin sensitivity, the approximated effective volume, and the corresponding concentration sensitivity.}   
	\label{oscillators_comp_lit}
\end{table*}

In Table \ref{oscillators_comp_lit}, we report a comparison of the performance of the oscillators studied in this work with respect to microwave oscillators based on normal metal resonators previously reported in literature \cite{Yalcin2008,matheoud2018,Schlecker2019,gualco2014,anders2012,matheoud2017}. The comparison table presents, for each device considered, the operating frequency, the operation temperature, the frequency noise, the estimated spin sensitivity, the estimated effective volume and the concentration sensitivity. The spin sensitivities recorded in this work are comparable, despite of the difference in operating frequency, to the values reported for most of the sensors operating in the 8 to 13 GHz range. Moreover, the estimated values of concentration sensitivity, also thanks to the bigger effective volume of our resonators, are comparable or better than the one reported in literature for sensors operating up to 150 GHz.

\section{OSC1: Estimation of $B_1$ from ESR linewidth measurements}					\label{section: ESR_600_MHz_sample_2}

In Fig.3 of the main text we reported measurements performed with a BDPA sample having a volume of about 1.2 nL placed at a distance of about 250 $\mu$m from the central line of the superconducting resonator in the oscillator OSC1.  In such measurements we observed negligible line broadening. Since BDPA has relaxation times  $T_1\cong T_2\cong100$ ns, the microwave field $B_1$ is less than 1 G at the sample position (see below). 
In Fig.\ref{figure: ESR_600_MHz_s2},  we report the results of experiments on a BDPA sample of  $120\times20\times20$ $\mu$m$^3$, corresponding to a volume of about 48 pL. The side of the sample is placed on the side of the central line of the resonator. All experiments, performed in the temperature range from 4.3 to 70 K, show line broadening from 5 to 10 G (see Fig\ref{figure: ESR_600_MHz_s2:a-b}).

\begin{figure*}[ht!]
	\centering 
	\includegraphics[width=\linewidth]{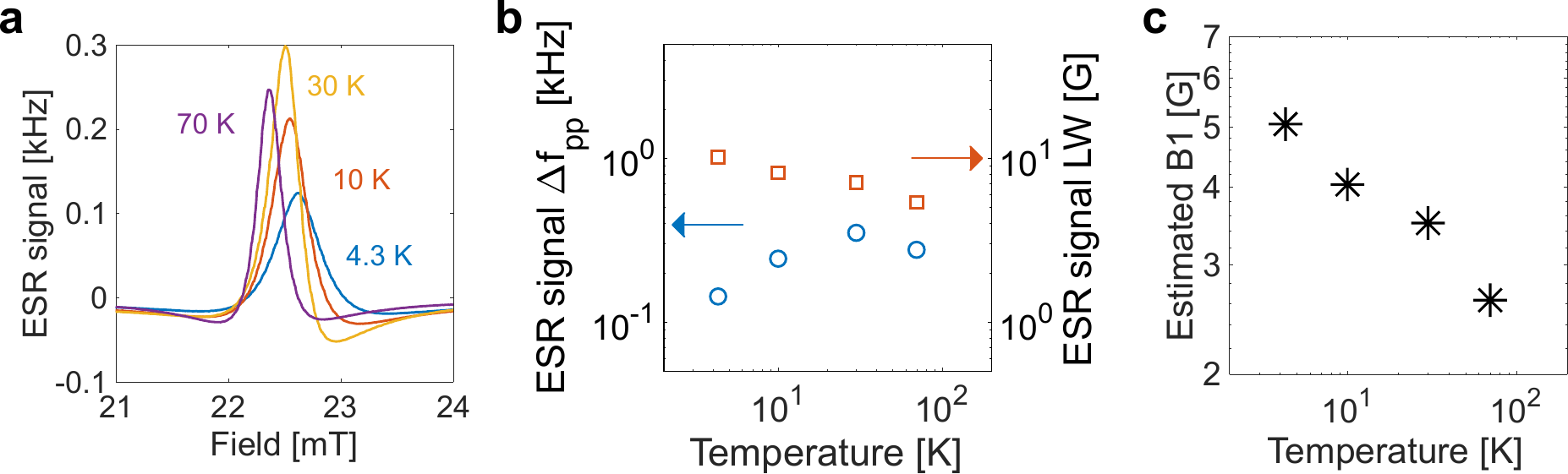} 
	\caption{Experiments to estimate $B_1$ from the line broadening, performed with a sample of BDPA having a volume of $120\times20\times20$  $\mu$m$^3$ placed at less then 1$\mu$m from the central line of the resonator of the 600 MHz oscillator (OSC1). \textbf{a} ESR spectra at different temperatures.\textbf{b} ESR signal amplitude vs temperature.  \textbf{c}  ESR signal linewidth vs temperature. \textbf{d} $B_1$ values vs temperature. All measurements are acquired with a field modulation amplitude of 0.4 G.}
	\label{figure: ESR_600_MHz_s2}
	\makeatletter
	\let\save@currentlabel\@currentlabel
	\edef\@currentlabel{\save@currentlabel a}\label{figure: ESR_600_MHz_s2:a}
	\edef\@currentlabel{\save@currentlabel b}\label{figure: ESR_600_MHz_s2:b}
	\edef\@currentlabel{\save@currentlabel c}\label{figure: ESR_600_MHz_s2:c}
	\edef\@currentlabel{\save@currentlabel a-10b}\label{figure: ESR_600_MHz_s2:a-b}
	\makeatother
\end{figure*}

For samples having homogeneously broadened lines the ESR linewidth at full width half maximum (FWHM) (in T) is given by
\begin{equation}
	\Delta B_{FWHM}=\frac{2}{T_2 \gamma_e}\sqrt{1+\gamma_e^2 B_1^2 T_1 T_2}
	\label{equation: delta_B}
\end{equation}

where $T_1$ is the longitudinal relaxation time (in s), $T_2$ is the transversal relaxation time (in s), $B_1$ is the microwave magnetic field (in T), and $\gamma_e\cong 2\pi\times 28$ GHz/T is the electron gyromagnetic ratio.
Hence, in presence of significant line broadening due the excessive $B_1$, the value of $B_1$ can be obtained from the measurement of the linewidth as
\begin{equation}
	B_1 = \sqrt{\frac{1}{\gamma_e^2 T_1 T_2} \bigg(\bigg(\frac{\Delta B_{FWHM}T_2 \gamma_e}{2}\bigg)^2 -1\bigg)}
	\label{equation: extraction_B1_from_LW}
\end{equation}
 
Consequently, the corresponding values of $B_1$ shown in Fig.\ref{figure: ESR_600_MHz_s2:c} are obtained from Eq. \ref{equation: extraction_B1_from_LW}.

From the extracted value of B$_1$ it is possible to roughly estimate the current flowing in the inductance line of the resonator in the operating conditions. As a first approximation, it is possible to consider the inductive line of the resonator as a infinitely thin planar wire. With this approximation, through a simple integration to extend the known result of Ampere's law, it is possible to estimate that the generated magnetic field is B$_{RF}$ according to Eq.\ref{equation: B_rf},

\begin{equation}
B_{RF} = \frac{ \mu_0 I}{2 w \pi}\log \bigg(\frac{r+w}{r}\bigg)
\label{equation: B_rf}
\end{equation}

where $w$ is the width of the line, $r$ the distance with respect to the side of the line and $I$ the current flowing into the line. From this equation, considering that $B_{RF}=2B_1$, it is possible to extrapolate an approximate value for the current flowing in the inductance line of the resonator in oscillation condition

\begin{equation}
I = \frac{4\pi B_1 w }{\mu_0 \log \bigg(\frac{r+w}{r}\bigg)}
\label{equation: I_rf}
\end{equation}

Hence, with $w=10$ $\mu$m, $B_1\cong0.5$ mT and $r = 10$ $\mu$m, we obtain $I\cong166$  mA. Considering a line thickness of 300 nm and a line width of 10 $\mu$m, the current density in the central line $J\cong6\times10^{10}$  A/m$^2$. As expected, this current density is significantly lower than the critical current density value $J_c\cong33\times10^{11}$ A/mm$^2$ at 4 K provided by the supplier of the YBCO film (Ceraco GmbH).  

Having determined an approximated value for the current flowing in the resonator, it is interesting to cross check if this value allows to confirm that the field produced at the location of the first sample (main text Fig.3) is indeed low enough to avoid the saturation of the BDPA sample. Assuming an approximate distance of 250 $\mu$m of the sample with respect to the line and using Eq. \ref{equation: B_rf}, we obtain $B_1=(1/2)B_{RF}\cong$ 0.2 G. With this $B_1$ value, the expected linewidth obtained from Eq. \ref{equation: delta_B} is 1.2 G, i.e., only marginally larger that the value of 1 G expected for $B_1\cong0$ and matching the linewidth measured at 4.3 K (see Fig.3e-3f).  

\section{Unitary field B1 spatial distribution}					\label{section: B1_spatial}

\begin{figure*}[ht!]
	\centering 
	\includegraphics[width=\linewidth]{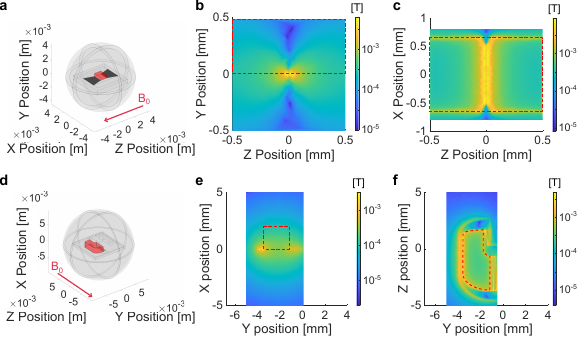} 
	\caption{B$_1$ unitary field (i.e., $\sqrt{B_{1x}^2 +B_{1y}^2}$ for a current of 1 A) for the resonators of OSC1, OSC2 and OSC3. \textbf{a}, \textbf{d}, Respectively for the resonator of OSC1 and for the resonator of OSC2 and OSC3. Geometry of the simulation with the resonator on top of the substrate (sapphire) in a sphere of air terminated by a perfectly matched layer (PML) shell. The red arrow shows the direction of the magnetic field B$_0$. The red volume represents the estimated effective volume for the resonators. \textbf{b}, B$_1$ unitary field on vertical section for OSC1 resonator. \textbf{c}, B$_1$ unitary field on horizontal section for OSC1 resonator. \textbf{e}, B$_1$ unitary field on vertical section for OSC2 and OSC3 resonator. \textbf{f}, B$_1$ unitary field on horizontal section for OSC2 and OSC3 resonator. The red dotted line in \textbf{b}, \textbf{c}, \textbf{e}, and \textbf{f} represent the boundaries of the considered effective volume on the planar sections respectively shown.}
	\label{figure: b1_spatial}
	\makeatletter
	\let\save@currentlabel\@currentlabel
	\edef\@currentlabel{\save@currentlabel a}\label{figure: b1_spatial:a}
	\edef\@currentlabel{\save@currentlabel b}\label{figure: b1_spatial:b}
	\edef\@currentlabel{\save@currentlabel c}\label{figure: b1_spatial:c}
	\edef\@currentlabel{\save@currentlabel d}\label{figure: b1_spatial:d}
	\edef\@currentlabel{\save@currentlabel e}\label{figure: b1_spatial:e}
	\edef\@currentlabel{\save@currentlabel f}\label{figure: b1_spatial:f}
	
	\makeatother
\end{figure*}

In Fig.\ref{figure: b1_spatial} we report on the spatial sensitivity of the two resonators used to build up the oscillators presented in this work. The simulations are performed with Comsol Multiphysics. More in details Fig.\ref{figure: b1_spatial:a} and Fig.\ref{figure: b1_spatial:d} present, respectively for the resonator of OSC1 and for the one of OSC2 and OSC3, the geometry of the simulation, showing the resonator on top of the Al2O3 monocrystalline substrate (sapphire) which is contained in a sphere of air terminated by a perfectly matched layer (PML) shell. The red volume in transparency represents the effective volume considered for the resonator (reported in Table I of the main text) and the red arrow shows the direction of the magnetic field B$_0$ along z axis. Fig.\ref{figure: b1_spatial:b} and Fig.\ref{figure: b1_spatial:e} show the plot for the B$_1$ unitary field on the vertical section, respectively for the resonator of OSC1 and for the one of OSC2 and OSC3. Fig.\ref{figure: b1_spatial:c} and Fig.\ref{figure: b1_spatial:f} show the plot for the B$_1$ unitary field on the horizontal. section, respectively for the resonator of OSC1 and for the one of OSC2 and OSC3. The red dotted lines in Fig.\ref{figure: b1_spatial:b}, Fig.\ref{figure: b1_spatial:c}, Fig.\ref{figure: b1_spatial:e} and Fig.\ref{figure: b1_spatial:f} represent the boundaries of the considered effective volume on the plane section shown in the corresponding figure. The effective volume is estimated as the volume resulting from the product of the resonator surface (the rectangular area around the central inductive line in case of the resonator of OSC1 and the area enclosed by the rounded rectangular coil and the capacitors in the case of OSC2 and OSC3) multiplied by an height approximated with the half-side of the resonator surface (see semitransparent red volume in Fig.\ref{figure: b1_spatial}). A similar approach, with correlation with measurements data extrapolated at different height from the coil surface, is also mentioned in Künstner et al. \cite{Kunstner2024} as a possible path for approximating the effective volume. Articles as Schlecker et al. \cite{Schlecker2019} take as approximation of the effective volume the cube of the coil diameter, considering also the half plane not effectively usable since hidden by the substrate. Other research articles use different approaches to estimate the effective volume. For instance, in Shtirberg et al. \cite{shtirberg2011}, the estimation of the effective volume is related to the ratio of the volume of a small hypothetical sample (for example, 1 $\mu$m$^3$) located at the point where the resonator’s microwave magnetic field is maximal, divided by the filling factor of this small sample. This approach could be maybe preferred if the chosen location of the sample is indeed the most sensitive on the resonator. We usually place our sample in a region of the resonator where the magnetic field is approximately corresponding to the average magnetic field in the effective volume shown as semi-transparent red volume in Fig.\ref{figure: b1_spatial}. For this reason, we believe our definition serves better the purpose in this case.

\section{OSC1: $Q$-factor of the resonator vs applied magnetic field}					\label{section: resonator_osc_01_vs_field}

\begin{figure*}[h!]
	\centering 
	\includegraphics[width=0.5\linewidth]{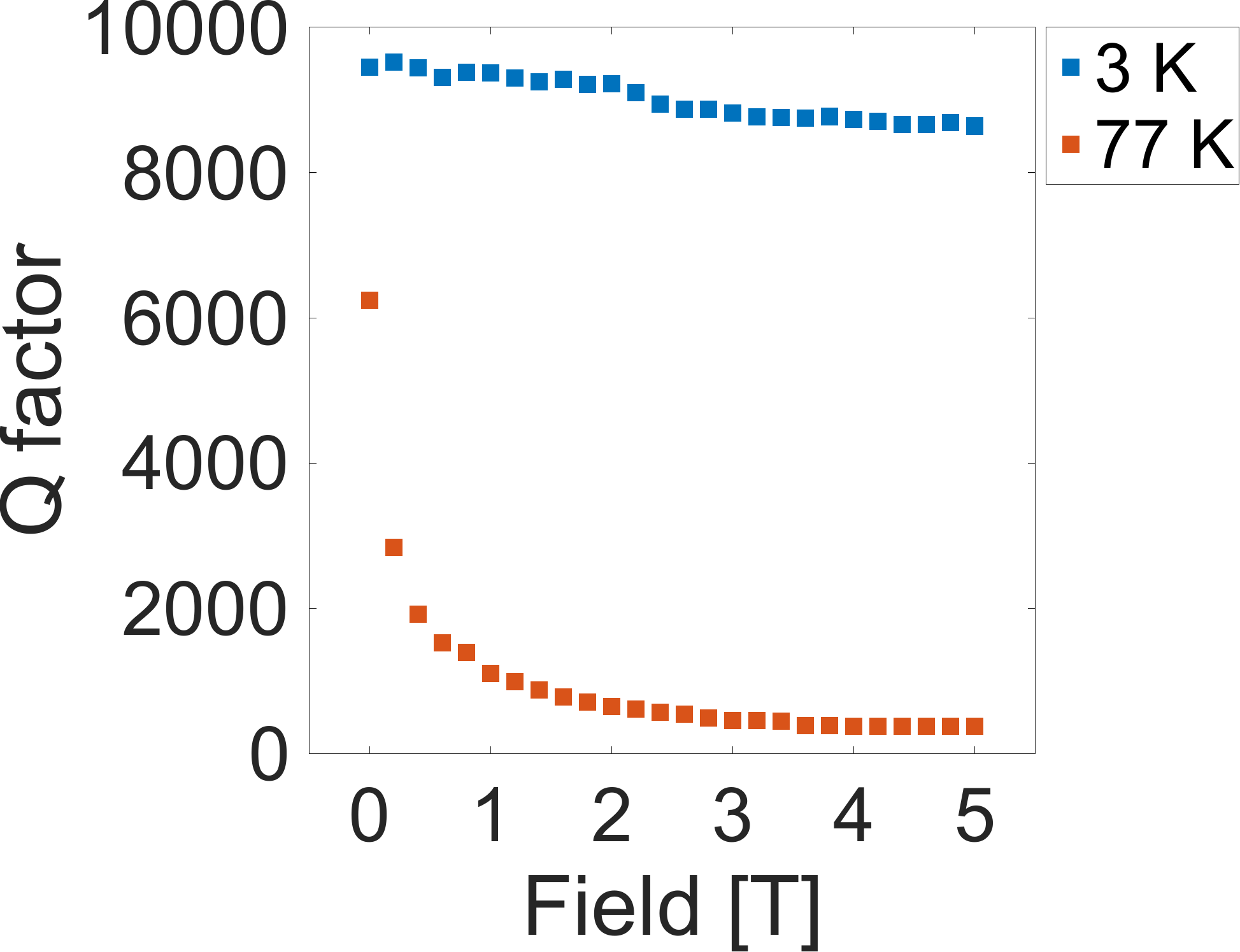}
	\caption{$Q$-factor of the resonator of the 0.6 GHz oscillator (OSC1) as a function of the applied magnetic field in the range from 0 to 5 T at 77 K and 3 K.}
	\label{figure: Q_factor_res_osc_01_vs_field_at_3K_and_77K}
\end{figure*}

The $Q$-factor of the resonator of the OSC1 is measured in the cryomagnet environment as a function of the externally applied magnetic field. In Fig.\ref{figure: Q_factor_res_osc_01_vs_field_at_3K_and_77K} are reported the results of these measurements. The magnetic field is applied parallel to the resonator plane in the range from 0 to 5 T. At 77 K the resonator $Q$-factor steeply drops from 6250 at 0 T to less than 400 at 5 T. On the other hand, at 3 K, the applied magnetic field determines only a minor reduction of the resonator $Q$-factor from 9450 at 0 T to 8640 at 5 T. This dependence of the resonator $Q$-factor on the applied static magnetic field explains the OSC1 behaviour with respect to the magnetic field described in \ref{section: freq_noise_vs_bias}.

\section{OSC1: Oscillation frequency and frequency noise vs bias voltage and magnetic field}					\label{section: freq_noise_vs_bias}

\begin{figure}[H]
	\centering 
	\includegraphics[width=\linewidth]{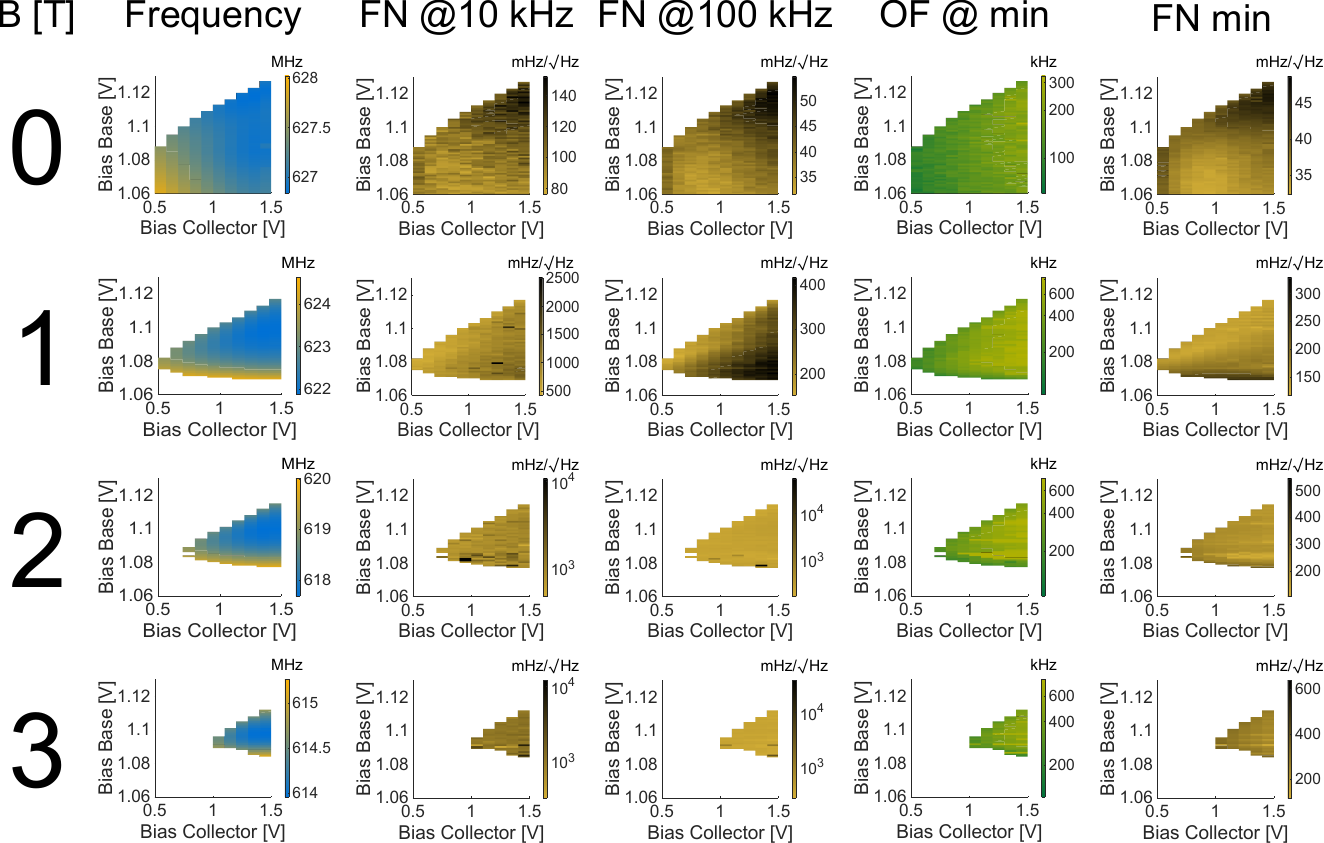} 
	\caption{0.6 GHz oscillator (OSC1): Oscillation frequency and frequency noise vs voltage bias and magnetic field at 77 K. }
	\label{figure: freq_noise_vs_bias_77K}
\end{figure}

\begin{figure}[H]
	\centering 
	\includegraphics[width=\linewidth]{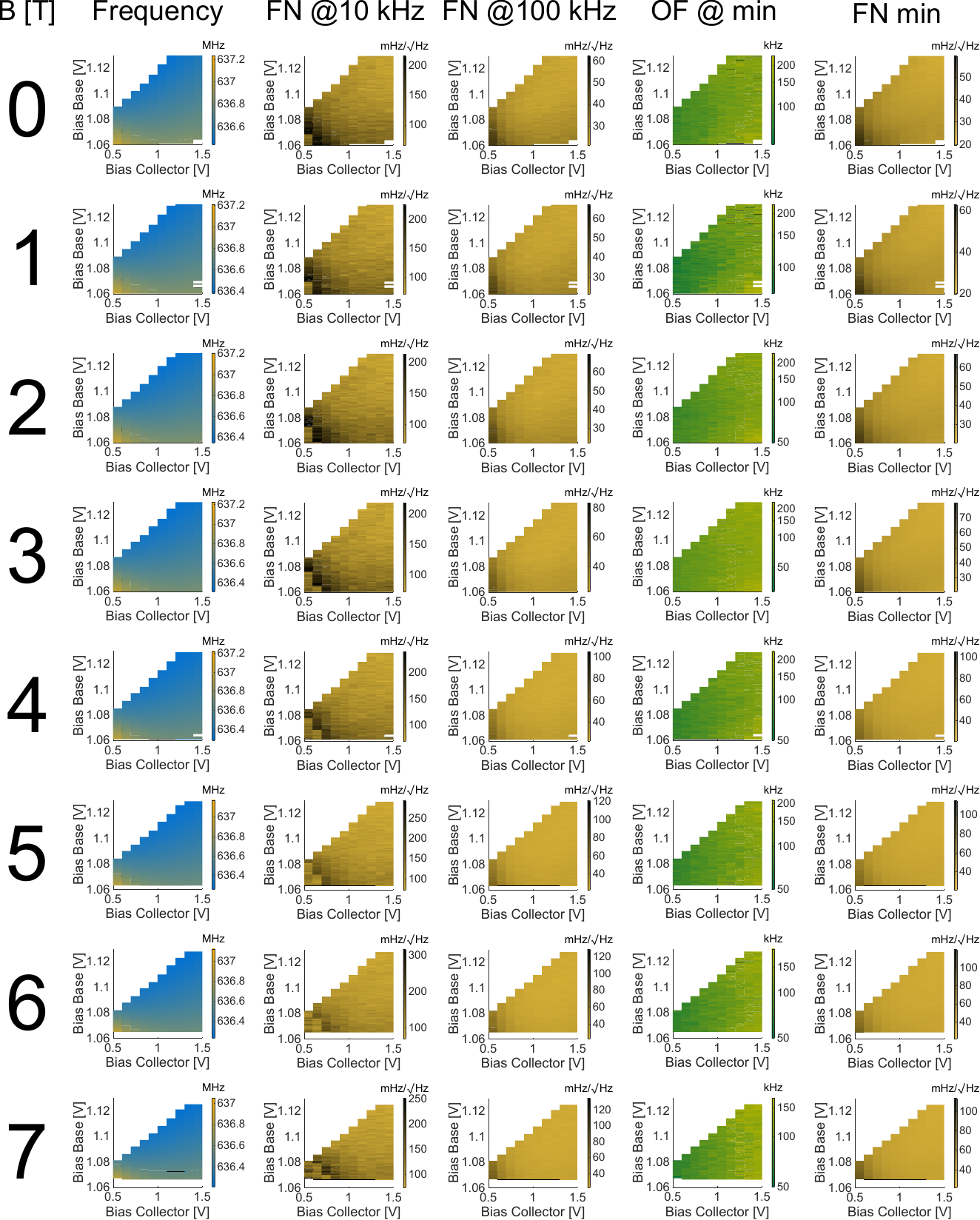} 
	\caption{0.6 GHz oscillator (OSC1): Oscillation frequency and frequency noise vs voltage bias and magnetic field at 3 K.}
	\label{figure: freq_noise_vs_bias_3K}
\end{figure}

In Fig.2 of the main text we reported measurements of the oscillation frequency and frequency noise of the 0.6 GHz oscillator (OSC1) vs bias voltages and magnetic field at 3 and 77 K. In such figure, we observe that the frequency noise steeply increases with the field at 77 K while only slightly increase with the field at 3 K. Additionally, the bias voltages working region for the oscillator is reduced with the increase of the magnetic field, an effect particularly pronounced at 77 K but almost absent at 3 K. In Fig.\ref{figure: freq_noise_vs_bias_77K} and Fig.\ref{figure: freq_noise_vs_bias_3K} we report the complete dataset of such experiments.


\bibliography{mybibfile}